\documentclass[a4paper,11pt]{article}
\pdfoutput=1
\usepackage{jheppub}
\usepackage{amsmath,amssymb,latexsym}
\usepackage{braket}
\usepackage{bm}
\usepackage{hepunits}
\usepackage{graphicx}
\usepackage[svgnames]{xcolor}
\usepackage{float}
\usepackage{mathtools}

\newcommand{\dif}{\mathrm{d}}

\definecolor{deepblue}{RGB}{0,0,180}

\allowdisplaybreaks

\title{Feynman Integral Reduction without Integration-By-Parts}
\author{Ziwen Wang,}
\author{Li Lin Yang}
\affiliation{Zhejiang Institute of Modern Physics, School of Physics, Zhejiang University, Hangzhou 310027, China}
\emailAdd{zwenwang@zju.edu.cn}
\emailAdd{yanglilin@zju.edu.cn}

\abstract{We present an interesting study of Feynman integral reduction that does not employ integration-by-parts identities. Our approach proceeds by studying the equivalence relations of integral contours in the Feynman parameterization. We find that the integration contour can take a more general form than that given by the Cheng-Wu theorem. We apply this idea to one-loop integrals, and derive universal reduction formulas that can be used to efficiently reduce any one-loop integral. We expect that this approach can be useful in the reduction of multi-loop integrals as well.}

\begin{document}

\maketitle

\section{Introduction}

Scattering amplitudes and Feynman integrals are core components of perturbative quantum field theories. Developing efficient computational methods is crucial for advancing cutting-edge phenomenological applications. To compute multi-loop scattering amplitudes, it is often necessary to handle a large number of complicated Feynman integrals. By exploiting the linear relations among these integrals, they can be expressed as linear combinations of a finite set of master integrals (MIs). This process is known as integral reduction, which significantly reduces the computational complexity. Integral reduction is also a key step in the method of differential equations~\cite{Kotikov:1990kg, Kotikov:1991pm, Remiddi:1997ny, Gehrmann:1999as} for evaluating the MIs.

Currently, the standard method for integral reduction is the Laporta algorithm~\cite{Laporta:2000dc} for solving the integration-by-parts (IBP) identities~\cite{Tkachov:1981wb, Chetyrkin:1981qh} of Feynman integrals. Many program packages implementing this algorithm are available, including \texttt{FIRE}~\cite{Smirnov:2023yhb}, \texttt{LiteRed}~\cite{Lee:2013mka}, \texttt{Reduze}~\cite{vonManteuffel:2012np} and \texttt{Kira}~\cite{Klappert:2020nbg}.
Traditionally, the IBP method has been developed mainly in the momentum representation. They can be formulated in other representations as well, such as in the Baikov representation \cite{Lee:2014tja, Larsen:2015ped, Bendle:2019csk, Chen:2022jux} or in the Feynman parameterization \cite{Lee:2014tja, Chen:2019mqc, Chen:2019fzm, Chen:2020wsh, Chen:2024xwt, Artico:2023jrc}.
For complicated multi-loop integrals, the system of IBP equations can become very large, and it is rather time-consuming or even impractical to solve them.
Recently, it has been proposed to generate a smaller set of IBP relations by using the algebraic geometry based method in \texttt{NeatIBP}~\cite{Wu:2023upw}, or by searching for a block-triangular system in \texttt{Blade}~\cite{Guan:2024byi}.
In addition to the IBP method, there exist other reduction techniques as well, e.g., Passarino-Veltman (PV) tensor reduction~\cite{Passarino:1978jh} and its improvements using auxiliary vectors~\cite{Feng:2021enk, Hu:2021nia, Feng:2022rwj, Feng:2022uqp, Feng:2022iuc, Feng:2022rfz, Li:2022cbx}, Ossola-Papadopoulos-Pittau (OPP) method~\cite{Ossola:2006us, Ossola:2007bb, Ellis:2007br}, unitarity cut method~\cite{Bern:1994zx, Bern:1994cg, Bern:1997sc, Britto:2004nc, Britto:2005ha, Britto:2006sj, Anastasiou:2006jv, Anastasiou:2006gt, Britto:2006fc, Britto:2007tt, Britto:2010um}, generating functions~\cite{Feng:2022hyg, Guan:2023avw, Hu:2023mgc, Li:2024rvo}, \textit{et al.}. We will not go into  details of these methods.

An alternative way to formulate the IBP relations is the so-called intersection theory~\cite{Mizera:2017rqa, Mastrolia:2018uzb, Frellesvig:2019kgj, Frellesvig:2019uqt, Mizera:2019vvs, Mizera:2020wdt, Frellesvig:2020qot, Caron-Huot:2021xqj, Caron-Huot:2021iev, Chestnov:2022alh, Fontana:2023amt, Brunello:2023rpq}, where a Feynman integral is regarded as a pairing between a differential form and an integration contour. The IBP relations are formulated as the equivalence relations among differential forms, which live in a so-called twisted cohomology group. The integral reduction can then be performed by calculating the intersection numbers between a pair of differential forms. Such an approach has been extensively developed in the Baikov representation~\cite{Baikov:1996iu, Lee:2010wea} of Feynman integrals, and recently has been developed for the Feynman parametrization as well~\cite{Lu:2024dsb}.

Within the framework of intersection theory, the equivalence relations can also be established among integration contours. The equivalence classes of contours form a twisted homology group. In principle, these equivalence relations can also be employed for integral reduction, but this approach has not been developed so far.

In this work, we initiate a study that exploits the equivalence of integration contours for integral reduction, based on the Feynman parameterization. One outcome of our study is an improvement of the Cheng-Wu theorem~\cite{Cheng:1969ab,Cheng:1987ga,Smirnov:2012gma}, such that the delta-function in the Feynman parameterization can be modified to a more general form. This essentially corresponds to modifying the integration contour.\footnote{Note that the deformation of integration contour in the Feynman parameterization is a common practice in the method of sector decomposition to avoid singularities in numerical integration \cite{Binoth:2005ff}. It is also demonstrated in \cite{Britto:2023rig} that generalized cuts can be implemented in the Feynman parameterization by appropriate choices of integral domains. These manipulations, however, has no connection with integral reduction.} We apply this to one-loop integrals and find that, by splitting the contour and further transforming each part of the contour, we can identify each part with a Feynman integral that is simpler than the original one. As a result, we can construct recursive reduction formulas purely by dealing with integration contours, without solving IBP relations. This approach does not generate any redundant information hidden in the IBP relations, and is therefore highly efficient.

The paper is organized as follows. In Section~\ref{sec:NewFeyPara}, we introduce the equivalence relations of integral contours in Feynman parameterization and use several simple examples to demonstrate our reduction method. In Section~\ref{sec:red-All}, we present our general method for one-loop integral reduction, and provide recursive formulas that can be easily implemented in computer algebra. In Section~\ref{sec:example}, we demonstrate our method through several examples, including a preliminary extension to higher loops. In Section~\ref{sec:summary}, we provide a summary and discuss the new challenges that may arise in future applications.

\section{Domain of integration and reduction of Feynman integrals}
\label{sec:NewFeyPara}

\subsection{Feynman parametrization and equivalence classes of integration domains}
\label{sec:NewCW}

An $L$-loop Feynman integral is defined by
\begin{equation}\label{eq:FeyInt}
    I(\bm{\nu}_n) = e^{\epsilon\gamma_E L} \int \frac{\dif^{d}k_1}{i\pi^{d/2}} \cdots \frac{\dif^{d}k_L}{i\pi^{d/2}} \, \frac{1}{D_1^{\nu_1} \cdots D_n^{\nu_n}} \,,
\end{equation}
where $\bm{\nu}_n \equiv \{ \nu_1,\cdots,\nu_n \}$, and $D_i$ are propagator denominators or irreducible scalar products. One can convert the above momentum representation into integrals over Feynman parameters. There are many variants of the Feynman parametrization, and one of them was introduced in \cite{Chen:2019mqc}. It takes the form (assuming all $\nu_j > 0$):
\begin{equation}\label{eq:ChenLP}
    I(\boldsymbol{\nu}_{n}) = C({\boldsymbol{\nu}}_{n}) \int_{0}^{\infty} \left( \prod_{j=0}^{n} \alpha_{j}^{\nu_{j}-1} \dif\alpha_{j} \right) \left( \alpha_{0} \, \mathcal{U} + \mathcal{F} \right)^{\lambda_{0}} \, \delta\left(1-\sum_{j\in\mathcal{S}}\alpha_{j}\right)\,,
\end{equation}
where $\mathcal{S}$ is a non-empty subset of $\{0,1,2,\cdots,n\}$, $\lambda_0 \equiv -d/2$, $\nu_{0}\equiv-\nu-(L+1)\lambda_0$, $\nu \equiv \sum_{j=1}^{n} \nu_j$, and the prefactor is given by
\begin{equation}
    C({\boldsymbol{\nu}}_{n}) \equiv {(-1)}^{\nu}e^{\epsilon\gamma_{E}L}\frac{\Gamma(-\lambda_{0})}{\prod_{i=0}^{n}\Gamma(\nu_{i})} \,.
\end{equation}
$\mathcal{U}$ and $\mathcal{F}$ are the so-called Symanzik polynomials. We denote
\begin{equation}
    \alpha_{1}D_{1}+\cdots+\alpha_{n}D_{n}\equiv\sum_{i,j=1}^{L}M_{ij}\,k_{i}\cdot k_{j}-2\sum_{i=1}^{L}k_{i}\cdot Q_{i}-J+i0 \, ,
\end{equation}
where $Q_i$ are combinations of external momenta. The two Symanzik polynomials can then be written as
\begin{equation}
    \mathcal{U} = \det(M) \, , \quad
    \mathcal{F} = \det(M) \left[ \sum_{i,j=1}^L M^{-1}_{ij} \, Q_i \cdot Q_j - J - i0 \right] \,  .
    \label{eq:uw_poly}
\end{equation}
From the above expressions, it is clear that $\mathcal{U}$ and $\mathcal{F}$ are homogeneous polynomials of degree $L$ and $L+1$ in the variables $\{\alpha_1,\cdots,\alpha_n\}$, respectively.

Note that the representation~\eqref{eq:ChenLP} can be applied to the cases where some $\nu_j \leq 0$ as well. For that we can introduce a regularization $\alpha_j^\rho$ into the integration measure. After performing the reduction, one takes the limit $\rho \to 0$ in the end. In the following, we will assume that such regulators are implicitly applied when necessary.

The fact that one can freely choose a subset of Feynman parameters appearing in the $\delta$-function in Eq.~\eqref{eq:ChenLP} follows from the so-called Cheng-Wu theorem~\cite{Cheng:1987ga}. We now note that Eq.~\eqref{eq:ChenLP} can actually be recasted into a more general form:
\begin{equation}\label{eq:newLP}
    I(\boldsymbol{\nu}_{n}) = C({\boldsymbol{\nu}}_{n}) \int_{0}^{\infty} \left( \prod_{j=0}^{n} \alpha_{j}^{\nu_{j}-1} \dif\alpha_{j} \right) \left( \alpha_{0} \, \mathcal{U} + \mathcal{F} \right)^{\lambda_{0}} \, \mathcal{X} \delta(\mathcal{X}-\mathcal{Y})\,,
\end{equation}
where $\mathcal{X}$ and $\mathcal{Y}$ are homogeneous functions of degree $0$ and $1$ in the variables ${\alpha_i}$, respectively. The integration domain is the hypersurface defined by the equation $\mathcal{X} - \mathcal{Y} = 0$ within the region $\mathbb{R}_+^{n+1}$ (where all $\alpha_j \geq 0$). The hypersurface should satisfy the condition that each ray going from the origin to $\mathbb{R}_+^{n+1}$ should have one and only one intersection point with the surface. For that purpose, $\mathcal{X} / \mathcal{Y}$ should be positive definite within $\mathbb{R}_+^{n+1}$. To reconcile the sign convention with Eq.~\eqref{eq:ChenLP}, we require that $\mathcal{X}$ and $\mathcal{Y}$ are both positive definite within $\mathbb{R}_+^{n+1}$. In practice, this usually requires that we work in a particular region of the kinematic variables. On the other hand, the reduction coefficients are rational functions and are correct in other kinematic regions as well. Therefore, the positivity constraint can be relaxed in the end.

It is easy to see that the original representation \eqref{eq:ChenLP} corresponds to the specific choice $\mathcal{X}=1$ and $\mathcal{Y}=\sum_{j\in\mathcal{S}}\alpha_{j}$. Note that by choosing $\mathcal{X}=1$ and $\mathcal{Y}=\alpha_0$, we can integrate out $\alpha_0$ to arrive at the so-called Lee-Pomeransky (LP) representation~\cite{Lee:2013hzt}:
\begin{equation}\label{eq:LP}
    I(\boldsymbol{\nu}_{n}) = C({\boldsymbol{\nu}}_{n}) \int_{0}^{\infty} \left( \prod_{j=1}^{n} \alpha_{j}^{\nu_{j}-1} \dif\alpha_{j} \right) \left( \mathcal{U} + \mathcal{F} \right)^{\lambda_{0}} \,.
\end{equation}
With these standard choices, the singularity structure of the integral within the parameter space can be analyzed using the Newton polytope associated with the polynomial $\alpha_0 \, \mathcal{U} + \mathcal{F}$. See, e.g., Refs.~\cite{Arkani-Hamed:2022cqe, Klausen:2023gui}. However, when $\mathcal{X}$ and $\mathcal{Y}$ are rational functions, the Newton polytope analysis needs to be modified to take them into account. We will not perform such analyses here, but simply note that it is always possible to find some regions of $\lambda_0$ where the integral converges.

It is straightforward to show that Eq.~\eqref{eq:newLP} with an arbitrary choice of $\mathcal{X}$ and $\mathcal{Y}$ (satisfying the conditions thereof) is equivalent to Eq.~\eqref{eq:ChenLP}. 
Starting from Eq.~\eqref{eq:newLP}, we insert the identity
\begin{equation}
    1=\int_{0}^{\infty}\delta\left(\eta-\sum_{j\in\mathcal{S}}\alpha_{j}\right)\,\dif\eta\,,
\end{equation}
and then perform the rescaling $\alpha_i\to \eta\,\alpha_i$. Using the homogeneity of the integrand as well as the functions $\mathcal{X}$ and $\mathcal{Y}$, Eq.~\eqref{eq:newLP} becomes
\begin{align}\label{eq:rescaling}
    I(\boldsymbol{\nu}_{n}) 
    =
    C({\boldsymbol{\nu}}_{n}) \int_{0}^{\infty} \left( \prod_{j=0}^{n} \alpha_{j}^{\nu_{j}-1} \dif\alpha_{j} \right) \left( \alpha_{0} \, \mathcal{U} + \mathcal{F} \right)^{\lambda_{0}} \, \delta\left(1-\sum_{j\in\mathcal{S}}\alpha_{j}\right)
    \int_0^\infty \dif \eta \, \mathcal{X} \delta(\eta \mathcal{X}-\mathcal{Y}) \,.
\end{align}
The positivity of $\mathcal{X}$ and $\mathcal{Y}$ ensures that the integration over $\eta$ equals unity. Therefore, Eq.~\eqref{eq:rescaling} is precisely Eq.~\eqref{eq:ChenLP}.

Eq.~\eqref{eq:newLP} with the freedom of choosing suitable $\mathcal{X}$ and $\mathcal{Y}$ functions will be our starting point to derive the reduction formulas. Note that we don't need IBP relations in the above derivation. On the other hand, our method is secretly equivalent to IBP reduction. In fact, the independence of Eq.~\eqref{eq:newLP} on $\mathcal{X}$ and $\mathcal{Y}$ can be associated with the Stoke's theorem, as we will demonstrate in the following.

We will follow an approach similar to that in Section 2.5.3 of~\cite{Weinzierl:2022eaz}. The $\delta$-function in Eq.~\eqref{eq:newLP} effectively restricts the integration onto the $n$-dimensional hypersurface $S_n$ embedded in $\mathbb{R}_+^{n+1}$, determined by $\mathcal{X}-\mathcal{Y}=0\,$:
\begin{equation}\label{eq:domain}
    S_{n} \equiv\left\{ \left(\alpha_{0},\cdots,\alpha_{n}\right)\in\mathbb{R}_+^{n+1} \big| \mathcal{X}-\mathcal{Y}=0 \right\} .
\end{equation}
Let's define an integrand function
\begin{equation}\label{eq:integrand}
    f \equiv C(\boldsymbol{\nu}_{n}) \left( \prod_{j=0}^{n}\alpha_{j}^{\nu_{j}-1} \right) \left( \alpha_{0} \, \mathcal{U} + \mathcal{F} \right)^{\lambda_{0}} \,,
\end{equation}
and an integration measure
\begin{equation}\label{eq:w}
    \omega \equiv \sum_{j=0}^{n} (-1)^{j} \, \alpha_{j}\,\dif\alpha_{0}\wedge\ldots\wedge\widehat{\dif\alpha_{j}}\wedge\ldots\wedge\dif\alpha_{n}\, ,
\end{equation}
where the hat indicates that the corresponding factor is omitted. When restricted to the surface $S_n$, $\omega$ can be parametrized by $n$ out of $n+1$ variables. Picking one of the variables (say, without loss of generality, $\alpha_0$) appearing in $\mathcal{X}-\mathcal{Y}\,$, we can express the integration measure $\omega$ on $S_n$ as
\begin{equation}
    \left.\omega\right|_{S_{n}} = - \left| \frac{\partial(\mathcal{X}-\mathcal{Y})}{\partial\alpha_{0}} \right|^{-1}_{S_{n}}
    \left[ \sum_{j=0}^n \alpha_j \frac{\partial(\mathcal{X}-\mathcal{Y})}{\partial\alpha_{j}} \right]_{S_{n}} \dif\alpha_{1}\wedge\ldots\wedge\dif\alpha_{n} \,,
\end{equation}
where the subscript $S_n$ indicates that $\alpha_0$ is replaced by a solution of $\mathcal{X}-\mathcal{Y}=0$. This expression is obtained by pulling back the integration measure on the hypersurface to the $n$-dimensional parameter space of a local coordinate chart. Note that the absolute value is necessary to enforce a consistent orientation convention. If the hypersurface cannot be covered by a single coordinate chart, the formula is understood with an implicit summation over the relevant local patches. Next, we use the fact that $\mathcal{X}$ and $\mathcal{Y}$ are homogeneous functions of degree $0$ and $1$, respectively. By Euler's homogeneous function theorem, we have $\sum_i \alpha_i \partial_{\alpha_i} \mathcal{X} = 0$ and $\sum_i \alpha_i \partial_{\alpha_i} \mathcal{Y} = \mathcal{Y}$, which imply
\begin{equation}
    \left[\sum_{j=0}^{n}\alpha_{j}\frac{\partial(\mathcal{X}-\mathcal{Y})}{\partial\alpha_{j}}\right]_{S_{n}}=\left.-\mathcal{Y}\right|_{S_{n}}=\left.-\mathcal{X}\right|_{S_{n}}\,.
\end{equation}
Combining the above information, we arrive at
\begin{align}
    \left.\omega\right|_{S_{n}} 
    &= \left| \frac{\partial(\mathcal{X}-\mathcal{Y})}{\partial\alpha_{0}} \right|_{S_{n}}^{-1}\left.\mathcal{X}\right|_{S_{n}}\dif\alpha_{1}\wedge\ldots\wedge\dif\alpha_{n} \nonumber \\
    &= \mathcal{X}\delta(\mathcal{X}-\mathcal{Y}) \, \dif\alpha_{0}\wedge\dif\alpha_{1}\wedge\ldots\wedge\dif\alpha_{n} \,,
\end{align}
Note that when solving for $\alpha_0$ from the $\delta$-function, one may obtain several solutions, which correspond to the different local patches on the hypersurface, as mentioned earlier. From the above expression, we see that the integral in Eq.~\eqref{eq:newLP} can be written as
\begin{equation}
    I(\boldsymbol{\nu}_{n}) = \int_{S_n} f \, \omega\,.
\end{equation}

\begin{figure}[ht!]
    \centering
    \includegraphics[width=0.3\linewidth]{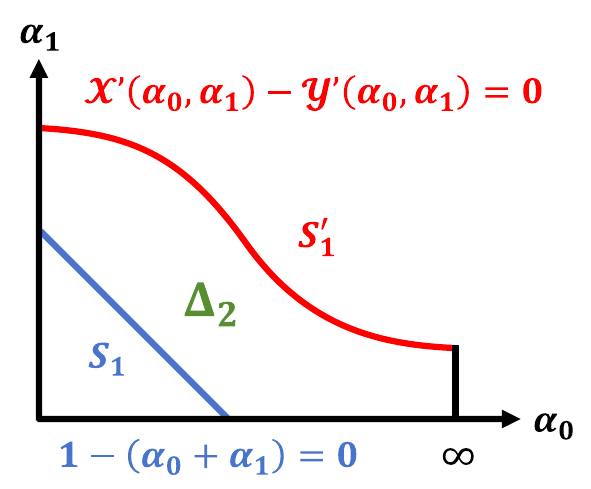}
    \caption{The independence of the integral on the integration domain for the case of two variables. $S_{1}$ is the integration domain defined by $\mathcal{X}=1$ and $\mathcal{Y}=\alpha_{0}+\alpha_{1}$, while $S_{1}'$ is the integration domain defined by two functions $\mathcal{X}'(\alpha_0,\alpha_1)$ and $\mathcal{Y}'(\alpha_0,\alpha_1)$.}
    \label{fig:domain}
\end{figure}

Suppose that $S_{n}$ and $S_{n}'$ are two $n$-dimensional oriented hypersurfaces defined by $\mathcal{X} -\mathcal{Y} = 0$ and $\mathcal{X}' -\mathcal{Y}' =0$, respectively. See Figure~\ref{fig:domain} for a simple illustration with $n=1$. We need to show that the integrals on $S_n$ and $S_{n}'$ are the same. For that we will employ Stokes' theorem. Let $\Delta_{n+1}$ be the region enclosed by $S_{n}$, $S_{n}'$, the coordinate hyperplanes defined by $\alpha_j = 0$ ($j=0,1,\ldots,n$), and possibly the hyperplanes at infinity. Note that $f$ is a homogeneous function of degree $(-n-1)$ in the variables $\{\alpha_j\}$. This can be used to demonstrate that the $n$-form $f \, \omega$ is closed, i.e., $\dif (f \,\omega) = \dif f \wedge \omega + f \, \dif \omega = 0$:
\begin{align}
    \dif f\wedge \omega &= \left(\sum_{j=0}^{n}\frac{\partial f}{\partial\alpha_{j}} \dif\alpha_{j}\right) \wedge\left(\sum_{j=0}^{n} (-1)^{j} \, \alpha_{j}\, \dif\alpha_{0}\wedge\ldots\wedge\widehat{\dif\alpha_{j}}\wedge\ldots\wedge\dif\alpha_{n}\right) \nonumber
    \\
    &= -(n+1) \, f \, \dif\alpha_{0}\wedge\dif\alpha_{1}\wedge\ldots\wedge\dif\alpha_{n} \nonumber
    \\
    &= -f \, \dif \omega\,.
\end{align}
Therefore, the integration of $\dif (f \,\omega)$ in any region $\Delta_{n+1}$ in $\mathbb{R}^{n+1}$ is zero. By Stokes' theorem, this means that
\begin{equation}\label{eq:stokes}
    \iint_{\Delta_{n+1}} \dif (f \,\omega) = \oint_{\partial\Delta_{n+1}} f \, \omega = 0\,.
\end{equation}

When restricted on the coordinate hyperplanes as part of $\partial\Delta_{n+1}$, $\omega$ always vanishes due to the factor of $\alpha_j$ in its definition \eqref{eq:w}. 
The boundary hyperplanes at infinity correspond to infinitesimally small solid angles. Therefore, if the integrand has no singularities on these boundaries, its integral is yields a vanishing contribution. If, on the other hand, the integrand exhibits singularities on the boundaries at infinity, one needs to introduce regulators to define the integral. In dimensional regularization, the integral is regarded as an analytic function of the complex variable $\lambda_0 = -d/2$. Unlike the integrals on $S_{n}$ and $S_{n}'$, the integrals at infinity are simpler to analyze since the functions $\mathcal{X}$ and $\mathcal{Y}$ no longer come into play. It is enough to realize that $\alpha_0 \, \mathcal{U} + \mathcal{F}$ is a homogeneous polynomial, while the variables cannot go to infinity uniformly when approaching the boundaries (otherwise the assumption below Eq.~\eqref{eq:newLP} would be violated). Therefore, it is always possible to find some region of $\lambda_0$ where $\alpha_0^{-(L+1) \lambda_0} (\alpha_0 \, \mathcal{U} + \mathcal{F})^{\lambda_0}$ goes to zero sufficiently fast: if $\alpha_0$ is finite at the boundary, we can choose $\lambda_0$ to be sufficiently negative; while if $\alpha_0 \to \infty$ at the boundary, we can choose $\lambda_0$ to be sufficiently positive.

The possible introduction of analytic regulators $\alpha_j^\rho$ does not change this conclusion, since $\rho$ should then be regarded as another complex variable on which the integral depends. Putting all the above together, we are now left with only $S_n$ and $S_n'$ in $\partial\Delta_{n+1}\,$.
After taking care of the orientations of the hypersurfaces, we finally arrive at
\begin{equation}
    \int_{S_{n}}f \, \omega = \int_{S_{n}'} f \, \omega \,.
    \label{eq:2-var-domains}
\end{equation}

In the above, we have shown that the two contours $S_n$ and $S'_n$ give rise to the same integral, and can be regarded as belonging to the same equivalence class. In the language of twisted homology (see, e.g., \cite{Aomoto:1414035} for more details), the equivalence classes of integration contours are elements (cycles) of a twisted homology group determined by the polynomial $\alpha_0 \, \mathcal{U} + \mathcal{F}$. This homology group is dual to the twisted cohomology group of the integrands. In the literature~\cite{Mizera:2017rqa, Mastrolia:2018uzb}, there have been extensive discussions on how to perform integral reduction using the vector-space structure of the cohomology groups. This can be done using the techniques of intersection theory~\cite{zbMATH00713739, zbMATH01270294, zbMATH02112802, zbMATH06267077, zbMATH06447521, zbMATH06502598, zbMATH06454357, zbMATH07531013, zbMATH07527773, zbMATH07733418}. The homology groups of integration contours also admit a vector-space structure, and in principle can be used for integral reduction as well. However, this path has not been followed in the literature to the best of our knowledge.

In the following, we will exploit the equivalence of integration domains to set up recursion relations that can be used to reduce one-loop integrals. We will write the relations in terms of index raising and lowering operators defined as:
\begin{align}
    \hat{j}^{+}I(\cdots,\nu_{j},\cdots) &\equiv \nu_{j}I(\cdots,\nu_{j}+1,\cdots) \,, \nonumber
    \\
    \hat{j}^{-}I(\cdots,\nu_{j},\cdots) &\equiv I(\cdots,\nu_{j}-1,\cdots) \,.
\end{align}
It is also useful to define an operator that set an index to zero:
\begin{equation}
    \hat{j}_{0}I(\cdots,\nu_{j},\cdots)\equiv I(\cdots,0,\cdots)\,.
\end{equation}
Before going into more general formalities, we first study a few simple examples.

\subsection{A reducible sector}
\label{sec:single}

\begin{figure}[th!]
    \centering
    \includegraphics[width=0.25\linewidth]{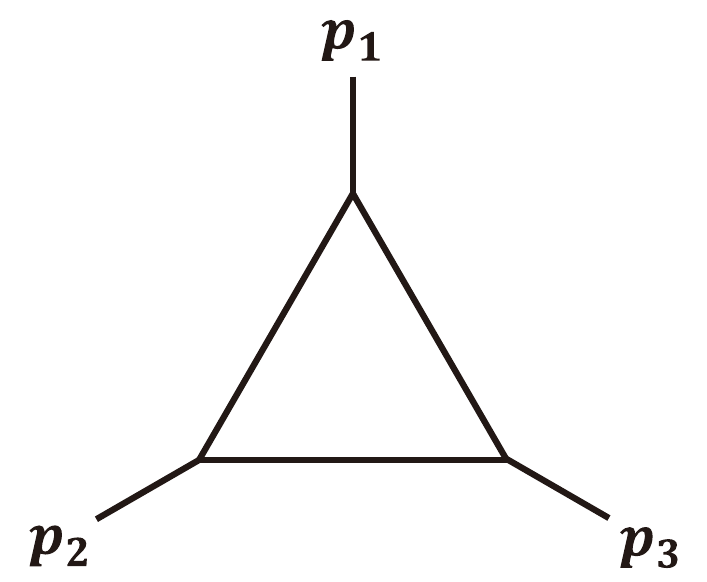}
    \caption{The massless triangle with $p_1^2 = s$ and $p_2^2=p_3^2=0$.}
    \label{fig:triangle}
\end{figure}

To illustrate our approach, we first consider a simple case where a sector is reducible to one of its sub-sectors. A specific example is the massless triangle diagram shown in Figure~\ref{fig:triangle}. The corner integral in this sector corresponding to $\bm{\nu}_3 = \bm{1}_3 = \{1,1,1\}$ is given by
\begin{equation}
    I({\boldsymbol{1}}_{3})=C({\boldsymbol{1}}_{3})\int_{0}^{\infty}\dif\alpha_{0}\dif\alpha_{1}\dif\alpha_{2}\dif\alpha_{3} \, \alpha_{0}^{d-4} \left[ \alpha_{0} \, \mathcal{U} + \mathcal{F} \right]^{-d/2} \mathcal{X} \, \delta(\mathcal{X}-\mathcal{Y}) \,,
\end{equation}
where $\mathcal{U}=\alpha_{1}+\alpha_{2}+\alpha_{3}$ and $\mathcal{F}=-\alpha_{1}\alpha_{2}s$. The above integral is reducible to a bubble integral. One can observe that $\alpha_3$ is present in $\mathcal{U}$ but absent in $\mathcal{F}$. We will show that there is a general rule: whenever a variable is present in $\mathcal{U}$ but absent in $\mathcal{F}$, the corresponding integral is reducible to the sub-sector without the variable.

Let's consider an $n$-point one-loop integral, and denote such a variable as $\alpha_{l}$. Defining $\mathcal{U}_{l,0} \equiv \mathcal{U}\big|_{\alpha_l \to 0}$, we can write $\mathcal{U} = \alpha_l + \mathcal{U}_{l,0}$. The corner integral in this sector can then be written as
\begin{equation}\label{eq:case-1lowertopo}
    I({\boldsymbol{1}}_{n})=C({\boldsymbol{1}}_{n}) \int_{0}^{\infty} \left( \prod_{j=0}^{n}\dif\alpha_{j} \right) \alpha_{0}^{d-n-1} \left[ \alpha_{0} \left(\alpha_{l}+\mathcal{U}_{l,0}\right)+ \mathcal{F}\right]^{\lambda_{0}} \mathcal{X} \, \delta(\mathcal{X}-\mathcal{Y})\,.
\end{equation}
We choose
\begin{equation}\label{eq:x-y_simple}
        \mathcal{X} =\frac{\mathcal{U}_{l,0}}{\alpha_{l}+\mathcal{U}_{l,0}} \,, \quad \mathcal{Y} = \alpha_{0} \,,
\end{equation}
and integrate out $\alpha_{0}$ using the $\delta$-function. After that, the variable $\alpha_l$ only appears in a power of $\alpha_{l}+\mathcal{U}_{l,0}$. The integration over $\alpha_l$ can then be performed using the formula
\begin{equation}
    \int_{0}^{\infty}\dif x\,x^{\beta-1}{\left(1+\frac{x}{\Lambda}\right)}^{\gamma-1} =
        \Lambda^{\beta} B(\beta,1-\beta-\gamma) \,, \quad (\Lambda > 0) \,,
\end{equation}
where the Beta function is
\begin{equation}
    B(z_1,z_2) = \frac{\Gamma(z_1) \, \Gamma(z_2)}{\Gamma(z_1+z_2)} \,.
\end{equation}

After integrating over $\alpha_l$, we arrive at
\begin{equation}
        I({\boldsymbol{1}}_{n})= C({\boldsymbol{1}}_{n}) \int_{0}^{\infty} \left( \prod_{j\neq0,l}\dif\alpha_{j} \right) \mathcal{U}_{l,0} \left( \mathcal{U}_{l,0} + \mathcal{F} \right)^{\lambda_{0}} \,.
\end{equation}
As promised, the integral now manifestly has the form of the LP representation in the sub-sector without $\alpha_l$. Using the index-changing operators, the above result can be written as
\begin{equation}\label{eq:decomp_sing}
    I({\boldsymbol{1}}_{n}) = \frac{1}{d-n-1}\sum_{k\neq l} \hat{k}^{+}\hat{l}_{0}I(\boldsymbol{1}_{n}) \,.
\end{equation}
Applying the above general formula to the triangle integral in Figure~\ref{fig:triangle}, we have
\begin{equation}
    I\left(1,1,1\right)=\frac{1}{d-4}\left[I\left(2,1,0\right)+I\left(1,2,0\right)\right] .
\end{equation}

\subsection{The bubble family}
\label{sec:bubble}

\begin{figure}[ht!]
    \centering
    \includegraphics[width=0.3\linewidth]{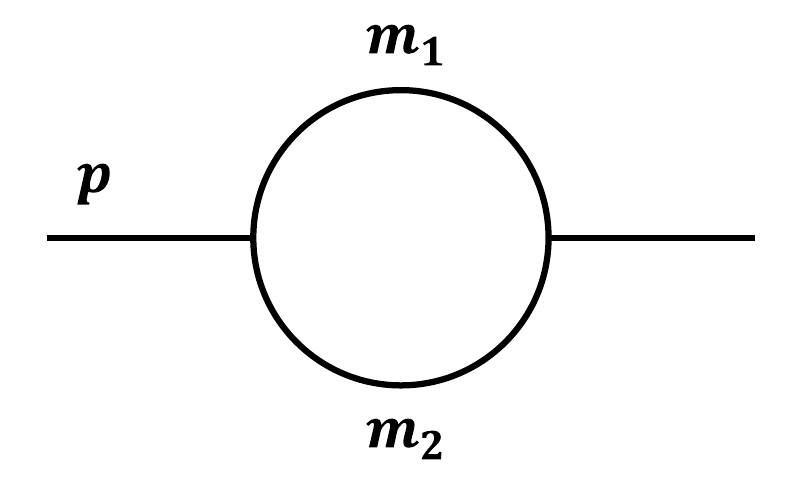}
    \caption{The bubble diagram with two masses.}
    \label{fig:bubble}
\end{figure}

We now turn to the reduction in an irreducible sector: the one-loop bubble with two masses shown in Figure~\ref{fig:bubble}. The kinematic variables are $m_{1}^{2}$, $m_{2}^{2}$ and $p^{2}=s$. The integrals in this family can be represented as 
\begin{multline}
    I(\nu_{1},\nu_{2})= C(\nu_{1},\nu_{2})\int_{0}^{\infty}\dif\alpha_{0}\dif\alpha_{1}\dif\alpha_{2}\, \alpha_{0}^{\nu_{0}-1}\alpha_{1}^{\nu_{1}-1}\alpha_{2}^{\nu_{2}-1}\mathcal{X}\delta(\mathcal{X}-\mathcal{Y})
    \\
    \times\left[(\alpha_{1}+\alpha_{2})\alpha_{0}+\alpha_{1}\alpha_{2}(m_{1}^{2}+m_{2}^{2}-s)+\alpha_{1}^{2}m_{1}^{2}+\alpha_{2}^{2}m_{2}^{2}\right]^{\lambda_{0}}\,.
    \label{eq:bubble-int}
\end{multline}

We consider the reduction of $I(1,2)$. Note that the integral shown in Eq.~\eqref{eq:bubble-int} for positive $\nu_1$ and $\nu_2$ does not possess the property of being reducible to sub-sectors as described in Section~\ref{sec:single}. Therefore, we need to split the integral into two parts: one part that is reducible to sub-sectors, and the other part that can be eventually related to $I(1,1)$. To this end, we consider the auxiliary integral 
\begin{equation}
    g(1,1)\equiv C(1,1)\int_{0}^{\infty}\dif\alpha_{0}\dif\alpha_{1}\dif\alpha_{2}\ \alpha_{0}^{d-3}\mathcal{X}\delta\left(\mathcal{X}-\mathcal{Y}\right)\left\{ \left[\alpha_{1}+(1-q_{1})\alpha_{2}\right]\alpha_{0}+\alpha_{1}^{2}m_{1}^{2}\right\} ^{\lambda_{0}}\,,
    \label{eq:g-bubble}
\end{equation}
where $q_{1}\equiv(m_{1}^{2}+m_{2}^{2}-s)/(2 m_{1}^{2})$.
The motivation for considering the above integral comes from two perspectives. First of all, the integral itself is reducible. According to the result in Section~\ref{sec:single}, one can see that $g(1,1)$ is reducible to the sub-sector without $\alpha_2$. Applying the method from Section~\ref{sec:single}, we obtain
\begin{equation}
        g(1,1)=  \frac{1}{d-3}\left(\frac{2m_{1}^{2}}{m_{1}^{2}-m_{2}^{2}+s}\right)I(2,0)\,.       
        \label{eq:g-I-bubble}
\end{equation}
On the other hand, the integral can be related to a linear combination of $I(1,1)$ and $I(1,2)$ through a change of variables. We perform the variable change $\alpha_{1}\to\alpha_{1}+q_{1}\alpha_{2}$ and $\alpha_{2}\to\alpha_{2}(1-W\alpha_{2}/\alpha_{0})$ in Eq.~\eqref{eq:g-bubble}, where $W\equiv\lambda(m_{1}^{2},m_{2}^{2},s)/\left[2(m_{1}^{2}-m_{2}^{2}+s)\right]$ and the Källén function is given by $\lambda(x,y,z)=x^{2}+y^{2}+z^{2}-2xy-2yz-2zx$. For convenience, we will work in the kinematic region where $W < 0$, such that the integration domain of $\alpha_2$ remains $[0,\infty]$ after the variable change. The reduction coefficients, being rational functions of kinematic variables, are insensitive to this choice. After the change of variables, we arrive at\footnote{It is easy to see that $\mathcal{X}$ and $\mathcal{Y}$ still satisfy the conditions below Eq.~\eqref{eq:newLP} in terms of the new variables. This is guaranteed since 1) the variable changes are homogeneous; and 2) $\mathcal{X}$ and $\mathcal{Y}$ are positive definite in the original integration domain, and must remain so in the transformed integral domain which maps bijectively to the original domain via the variable changes. The same consideration applies to variable changes employed later.}
\begin{align}\label{eq:g-l-bubble}
    g(1,1) &= C(1,1)\int_{0}^{\infty}\dif\alpha_{0}\dif\alpha_{2}\int_{-q_{1}\alpha_{2}}^{\infty}\dif\alpha_{1}\,\alpha_{0}^{d-3}\mathcal{X}\delta(\mathcal{X}-\mathcal{Y}) \left(1-2W\frac{\alpha_{2}}{\alpha_{0}}\right) \nonumber
    \\
    &\quad \times\left[(\alpha_{1}+\alpha_{2})\alpha_{0}+\alpha_{1}\alpha_{2}(m_{1}^{2}+m_{2}^{2}-s)+\alpha_{1}^{2}m_{1}^{2}+\alpha_{2}^{2}m_{2}^{2}\right]^{\lambda_{0}} \nonumber
    \\
    &= I(1,1) + \frac{2W}{d-3} \, I(1,2) + h_1(1,1) \,,
\end{align}
where we have split the integration range of $\alpha_1$ and define
\begin{align}
    h_1(1,1) &\equiv C(1,1)\int_{0}^{\infty}\dif\alpha_{0}\dif\alpha_{2}\int_{-q_{1}\alpha_{2}}^{0}\dif\alpha_{1}\,\alpha_{0}^{d-3}\mathcal{X}\delta(\mathcal{X}-\mathcal{Y}) \left(1-2W\frac{\alpha_{2}}{\alpha_{0}}\right) \nonumber
    \\
    &\quad \times\left[(\alpha_{1}+\alpha_{2})\alpha_{0}+\alpha_{1}\alpha_{2}(m_{1}^{2}+m_{2}^{2}-s)+\alpha_{1}^{2}m_{1}^{2}+\alpha_{2}^{2}m_{2}^{2}\right]^{\lambda_{0}} \,.
\end{align}
The $h_1(1,1)$ function is also reducible to sub-sectors. This can be seen by further applying the variable change $\alpha_{0}\to\alpha_{0}\alpha_{2}/(\alpha_{2}+\alpha_{1}/q_{1})$, $\alpha_{1} \to -\alpha_{1}\alpha_{2}/(\alpha_{2}+\alpha_{1}/q_{1})$. Denoting $\alpha_{1}'\equiv\alpha_{1}\left[1-W(2\alpha_{2}+\alpha_{1}/q_{1})/\alpha_{0}\right]$, we have
\begin{align}\label{eq:h-I-bubble}
    h_1(1,1) &= C(1,1)\int_{0}^{\infty}\dif\alpha_{0}\dif\alpha_{1}'\dif\alpha_{2}\,\alpha_{0}^{d-3}\mathcal{X}\delta(\mathcal{X}-\mathcal{Y})\left\{ \left[(1-q_{1})\frac{\alpha_{1}'}{q_{1}}+\alpha_{2}\right]\alpha_{0}+\alpha_{2}^{2}m_{2}^{2}\right\} ^{\lambda_{0}} \nonumber
    \\
    &= \frac{1}{d-3} \, \frac{m_{1}^{2}+m_{2}^{2}-s}{m_{1}^{2}-m_{2}^{2}+s} \, I(0,2)\,,
\end{align}
where we have again utilized the method of Section~\ref{sec:single} (assuming $0 < q_1 < 1$).
Combining Eq.~\eqref{eq:g-I-bubble}, \eqref{eq:g-l-bubble} and \eqref{eq:h-I-bubble}, we arrive at
\begin{equation}
    I(1,2)=(3-d) \, \frac{m_{1}^{2}-m_{2}^{2}+s}{\lambda(m_{1}^{2},m_{2}^{2},s)} \, I(1,1)+\frac{2 m_{1}^{2}}{\lambda(m_{1}^{2},m_{2}^{2},s)} \, I(2,0)-\frac{m_{1}^{2}+m_{2}^{2}-s}{\lambda(m_{1}^{2},m_{2}^{2},s)} \, I(0,2) \,,
\end{equation}
which precisely agrees with the result of IBP reduction.

From the above examples, one may find that our procedure mainly consists of two steps: 1) transform of the integration contours either by explicit choices of the $\mathcal{X}$ and $\mathcal{Y}$ functions, or by appropriate variable changes; 2) split the integration contour into several parts, and identify each part manifestly as a Feynman integral. One may then wonder how these contour transforms are constructed. In the next Section, we will present the general method and the explicit recursive reduction formula for one-loop integrals.

\section{The general method for one-loop integral reduction}\label{sec:red-All}

Based on the examples in the previous Section, we now describe our reduction approach for generic one-loop integrals. The final outcome of this Section is a system of relations that express a more complicated integral (with higher powers of propagators and/or higher ranks of numerators) as a linear combination of simpler integrals (with lower powers of propagators and/or lower ranks of numerators). By iteratively applying these relations, it gives a shortcut to reduce a integral to corner integrals with indices being either 1 or 0.

In essence, our iterative relations are equivalent to a symbolic solution to IBP equations, which at one-loop can be obtained through other methods as well (see, e.g., \cite{Feng:2022rwj} and references therein). In the language of linear systems, these iterative relations automatically generate a system in an upper triangular form (row echelon form). In other words, we have effectively performed the forward elimination in the Gaussian elimination method of solving linear systems with no cost, and applying the iterative relations is equivalent to the step of back substitution. It is well-known that the computational complexity of Gaussian elimination is dominated by that of forward elimination, which scales as $N^3$ where $N$ is the number of equations. On the other hand, the complexity of back substitution only scales as $N^2$. Therefore, it is natural to expect that our method provides better efficiency than the traditional IBP methods.

\subsection{More powers v.s. more variables}
\label{sec:Decrease}

The goal of one-loop integral reduction is to reduce the indices $\nu_j$ to either $0$ or $1$. We now introduce an interesting technique to lower an index by one, at the cost of adding an auxiliary integration variable. While this seems to be meaningless at first sight, it will be employed in the derivation of the final reduction formula. 

For an arbitrary function $g(u)$, we can derive the following integral relation:
\begin{align}
    \label{eq:decrease_index}
    \frac{1}{\Gamma(n+1)}\int_{0}^{\infty}\dif u\,u^{n}g(u) & =\frac{1}{\Gamma(n+1)}\int_{0}^{\infty}\dif u\left[n\int_{0}^{u}\dif x\,x^{n-1}\right]g(u)\nonumber\\
    & =\frac{1}{\Gamma(n)}\int_{0}^{\infty}\dif x\int_{x}^{\infty}\dif u\left[x^{n-1}g(u)\right]\nonumber\\
    & =\frac{1}{\Gamma(n)\,\Gamma(1)}\int_{0}^{\infty}\dif x\int_{0}^{\infty}\dif y\left[x^{n-1}g(x+y)\right].
\end{align}   
Here, we emphasize that a regularization $n \to n + \rho$ is implicitly assumed if $n \leq 0$. We may apply the above relation to the Feynman parametric integrals. We define
\begin{multline}\label{eq:de_power}
    I(\nu_{1},\cdots,\left[\nu_{l},1\right],\cdots,\nu_{n})
    \\
    \equiv C({\boldsymbol{\nu}}_{n},1) \int_{0}^{\infty} \dif\beta_{l} \int_{0}^{\infty} \left( \prod_{j=0}^{n} \alpha_{j}^{\nu_{j}-1} \dif\alpha_{j} \right) \mathcal{X} \delta(\mathcal{X}-\mathcal{Y}) \left[\left({\alpha_{0}\,\mathcal{U}}+{\mathcal{F}}\right)\big|_{\alpha_{l}\to\alpha_{l}+\beta_{l}}\right]^{\lambda_{0}}\,.
\end{multline}
A few words are needed to explain a subtlety in the above definition. Both the prefactor $C$ and the power of $\alpha_0$ in the integrand involves $\nu$, which is defined as the sum of all indices in the argument of $I$ and $C$, as can be seen below Eq.~\eqref{eq:ChenLP}. It then follows that, in the above expression, we have $\nu = 1 + \sum_{j=1}^n \nu_j$. Note that the value of $\nu_0 \equiv -\nu - (L+1)\lambda_0$ is determined accordingly. Similar considerations apply for the functions $g(\bm{\nu})$ and $h_i(\bm{\nu})$ that will be introduced later.

Applying Eq.~\eqref{eq:decrease_index} to Eq.~\eqref{eq:de_power}, we obtain
\begin{equation}\label{eq:de_power_2}
    I(\nu_{1},\cdots,\nu_{l}+1,\cdots,\nu_{n})=I(\nu_{1},\cdots,\left[\nu_{l},1\right],\cdots,\nu_{n})\,.
\end{equation}
From a different point of view, the above relation can also be obtained by splitting $D_l^{-\nu_l-1}$ as $D_l^{-\nu_l} \, D_l^{-1}$, and introducing two Feynman parameters $\alpha_l$ and $\beta_l$ for the two factors.
This relation will play a crucial role in subsequent derivations. It is worth noting that it also holds for general $L$-loop integrals.

\subsection{Reduction for irreducible sectors}\label{sec:red-irred-sectors}

We now consider a general one-loop integral in the Feynman representation:
\begin{equation}\label{eq:general-I}
    I({\boldsymbol{\nu}}_{n})=C({\boldsymbol{\nu}}_{n})\int_{0}^{\infty} \left( \prod_{j=0}^{n} \alpha_{j}^{\nu_{j}-1} \dif\alpha_{j} \right) \mathcal{X} \delta(\mathcal{X}-\mathcal{Y}) \left[\left(\boldsymbol{1}^{T}\boldsymbol{\alpha}\right)\alpha_{0}+\boldsymbol{\alpha}^{T}\boldsymbol{Z}\boldsymbol{\alpha}\right]^{\lambda_{0}}\,,
\end{equation}
where $\boldsymbol{1}$ denotes a column vector of length $n$ with all elements equal to $1$, and $\boldsymbol{\alpha} \equiv (\alpha_1,\cdots,\alpha_n)^{T}$. It is evident that
\begin{equation}
    \mathcal{U} = \boldsymbol{1}^{T}\boldsymbol{\alpha} \,, \quad \mathcal{F} = \boldsymbol{\alpha}^{T}\boldsymbol{Z}\boldsymbol{\alpha} \,,
\end{equation}
where the Gram matrix $\boldsymbol{Z}$ is symmetric. For an irreducible sector, we have $\det(\bm{Z}) \neq 0$ (but the reverse is not true, as we will see in the ``magic relations'' discussed later).

In the following, we will aim for decreasing the power $\nu_l > 1$ of the variable $\alpha_l$, where $l \in \{1,\cdots,n\}$. For that we will need to study the submatrix of $\boldsymbol{Z}$ obtained by removing the $l$-th row and $l$-th column. We will denote this submatrix as $\boldsymbol{Z}(\hat{l},\hat{l})$, and refer to it as ``the $(l,l)$-submatrix''. A closely related concept is the $(i,j)$-minor of $\boldsymbol{Z}$, which is the determinant of the submatrix of $\boldsymbol{Z}$ without the $i$-th row and $j$-th column. The $(i,j)$-cofactor is further defined by multiplying the $(i,j)$-minor with $(-1)^{i+j}$. We will denote the $(i,j)$-cofactor of $\bm{Z}$ as $Z_{i,j}$.

The $(l,l)$-submatrix can be either non-singular with rank $n-1$, or singular with a smaller rank. In the following we will discuss the two situations separately.

\subsubsection{When the \texorpdfstring{$(l,l)$}{}-submatrix is non-singular}\label{sec:non-sing}

Similar to the idea of Eq.~\eqref{eq:g-bubble}\footnote{As in Section~\ref{sec:bubble}, the kinematic invariants appearing in the derivation of this section are implicitly assumed to lie in an appropriate domain.}, we introduce an auxiliary integral 
\begin{multline}\label{eq:g-def}
    g({\boldsymbol{\nu}}_{n},{\boldsymbol{1}}_{n})\equiv C({\boldsymbol{\nu}}_{n},{\boldsymbol{1}}_{n})\int_{0}^{\infty}\left(\prod_{j=1}^{n}{\dif\beta_{j}}\right)\int_{0}^{\infty}\left(\prod_{j=0}^{n}\alpha_{j}^{\nu_{j}-1}\,\dif\alpha_{j}\right)\,\mathcal{X}\delta(\mathcal{X}-\mathcal{Y})
    \\
    \times \left\{ \left[\boldsymbol{1}^{T}\left(\boldsymbol{{\alpha}}+{\boldsymbol{\beta}}^{\left(l\right)}-\boldsymbol{q}\beta_{l}\right)\right]\alpha_{0}+{\left(\boldsymbol{\alpha}+{\boldsymbol{\beta}}^{\left(l\right)}\right)}^{T}\boldsymbol{Z}\left(\boldsymbol{\alpha}+{\boldsymbol{\beta}}^{\left(l\right)}\right) \right\}^{\lambda_{0}}\,,
\end{multline}
where $\boldsymbol{\beta}^{\left(l\right)}$ denotes the vector obtained by replacing the $l$-th element (i.e., $\beta_l$) of the vector $\boldsymbol{\beta} \equiv (\beta_1,\cdots,\beta_n)^{T}$ with $0$, and the elements of the vector $\bm{q}$ are given by
\begin{equation}\label{eq:q_define}
    q_{k}\equiv-\frac{Z_{k,l}}{Z_{l,l}}\,,
\end{equation}
where we recall that $Z_{i,j}$ is the $(i,j)$-cofactor of $\bm{Z}$.

Using the method from Section~\ref{sec:single}, we can extract $\beta_{l}$ and integrate it out with an appropriate choice of $\mathcal{X}$ and $\mathcal{Y}$. We arrive at
\begin{align}\label{eq:g-expr}
    g({\boldsymbol{\nu}}_{n},{\boldsymbol{1}}_{n}) =
    \frac{1}{\nu_{0}-1}\left(-\frac{1}{q}\right)\sum_{j=1}^{n}\left[\hat{j}^{+}\hat{l}^{-}I({\boldsymbol{\nu}}_{n}+{\boldsymbol{1}}_{n})\right]\,,
\end{align}   
where $q\equiv\sum_{k=1}^{n} q_{k}$, and we have used Eq.~\eqref{eq:de_power_2} to absorb the extra $\beta_j$ variables at the cost of increasing the powers of $\alpha_j$. Note that $\nu_0$ is defined in terms of the sequence of indices $(\bm{\nu}_n,\bm{1}_n)$, as explained below Eq.~\eqref{eq:de_power}.

On the other hand, Eq.~\eqref{eq:g-def} can be transformed in another way with the variable change
\begin{equation}
    \beta_{j}\to\beta_{j}+q_{j}\beta_{l}\,,\; \left(\forall j\neq l\right),\quad\beta_{l}\to\beta_{l}\left(1-W\,\frac{\beta_{l}+2\alpha_{l}}{\alpha_{0}}\right),
\end{equation}
where 
\begin{equation}
    W\equiv-\frac{\det(\boldsymbol{Z})}{\sum_{j=1}^{n}{Z_{j,l}}} = \frac{\det(\boldsymbol{Z})}{q \, Z_{l,l}} \,.
\end{equation}
This leads to
\begin{multline}
    g({\boldsymbol{\nu}}_{n},{\boldsymbol{1}}_{n})=C({\boldsymbol{\nu}}_{n},{\boldsymbol{1}}_{n})\int_{0}^{\infty}\dif\beta_{l}\left(\prod_{j\neq l}\int_{-q_{j}\beta_{l}}^{\infty}{\dif\beta_{j}}\right)\int_{0}^{\infty}\left(\prod_{j=0}^{n}\alpha_{j}^{\nu_{j}-1}\,\dif\alpha_{j}\right)\,\mathcal{X}\delta(\mathcal{X}-\mathcal{Y})\\
    \times\left(1-2W\frac{\alpha_{l}+\beta_{l}}{\alpha_{0}}\right)\left[\boldsymbol{1}^{T}\left(\boldsymbol{\alpha}+\boldsymbol{\beta}\right)\alpha_{0}+{\left(\boldsymbol{\alpha}+\boldsymbol{\beta}\right)}^{T}\boldsymbol{Z}{\left(\boldsymbol{\alpha}+\boldsymbol{\beta}\right)}\right]^{\lambda_{0}}\,.
\end{multline}    
We can now split the integration domain of $\beta_j\,$. First, we introduce the contour-splitting identity
\begin{equation}\label{eq:integral_equal}
    \prod_{j\neq l}{\int_{-x_{l}}^{\infty}\dif x_{j}}=\prod_{j\neq l}{\int_{0}^{\infty}\dif x_{j}}+\sum_{k\neq l}\int_{-x_{l}}^{0}\dif x_{k}\prod_{j\neq l,k}{\int_{x_{k}}^{\infty}\dif x_{j}} \,.
\end{equation}
\begin{figure}[H]
    \centering
    \includegraphics[width=0.3\linewidth]{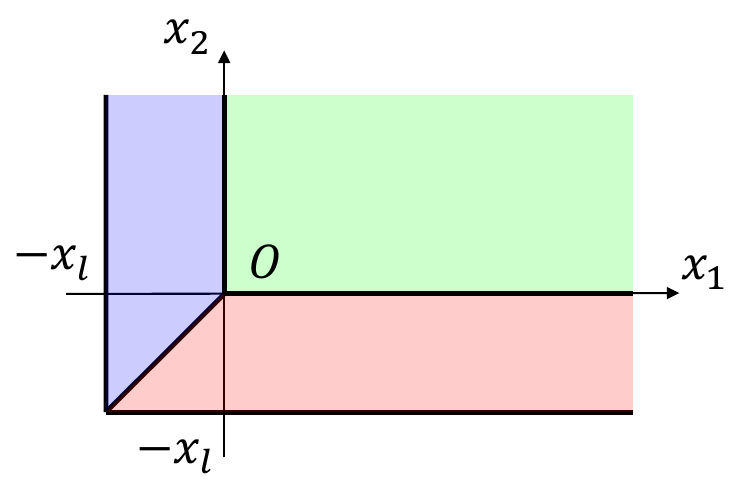}
    \caption{The contour-splitting in two dimensions}
    \label{fig:contour-splitting}
\end{figure}
whose geometric meaning is natural and straightforward. In two dimensions, it corresponds to the domain decomposition illustrated in Figure~\ref{fig:contour-splitting}. After a rescaling of the variables, we arrive at
\begin{equation}\label{eq:g-I-h}
    g({\boldsymbol{\nu}}_{n},{\boldsymbol{1}}_{n})=I({\boldsymbol{\nu}}_{n}+{\boldsymbol{1}}_{n})+\frac{2W}{\nu_{0}-1}\hat{l}^{+}I({\boldsymbol{\nu}}_{n}+{\boldsymbol{1}}_{n})+\sum_{k\neq l}h_{k}({\boldsymbol{\nu}}_{n},{\boldsymbol{1}}_{n})\,,
\end{equation}
where we have defined
\begin{multline}
    h_{k}({\boldsymbol{\nu}}_{n},{\boldsymbol{1}}_{n})\equiv C({\boldsymbol{\nu}}_{n},{\boldsymbol{1}}_{n})\int_{0}^{\infty}\dif\beta_{l}\int_{-q_{k}\beta_{l}}^{0}\dif\beta_{k}\left(\prod_{j\neq l,k}{\int_{q_{j}\beta_{k}/q_{k}}^{\infty}\dif\beta_{j}}\right)\int_{0}^{\infty}\left(\prod_{j=0}^{n}\alpha_{j}^{\nu_{j}-1}\,\dif\alpha_{j}\right)\\
    \times\left(1-2W\frac{\alpha_{l}+\beta_{l}}{\alpha_{0}}\right)\left[\boldsymbol{1}^{T}\left(\boldsymbol{\alpha}+\boldsymbol{\beta}\right)\alpha_{0}+{\left(\boldsymbol{\alpha}+\boldsymbol{\beta}\right)}^{T}\boldsymbol{Z}{\left(\boldsymbol{\alpha}+\boldsymbol{\beta}\right)}\right]^{\lambda_{0}}\,\mathcal{X}\delta(\mathcal{X}-\mathcal{Y})\,.
\end{multline}    

Our next task is to transform $h_{k}({\boldsymbol{\nu}}_{n})$ into Feynman integrals. For that we perform the variable change
\begin{equation}
    \alpha_{j}\to\alpha_{j}\frac{\beta_{l}}{\beta_{l}+\frac{\beta_{k}}{q_{k}}}\,,\; \left(\forall j\right), \quad \beta_{k}\to-\beta_{k}\frac{\beta_{l}}{\beta_{l}+\frac{\beta_{k}}{q_{k}}}\,, \quad \beta_{j}\to\left(\beta_{j}-q_{j}\frac{\beta_{k}}{q_{k}}\right)\frac{\beta_{l}}{\beta_{l}+\frac{\beta_{k}}{q_{k}}}\,,\ \left(\forall j\neq l,k\right) .    
\end{equation}
We can then write
\begin{multline}
    h_{k}({\boldsymbol{\nu}}_{n},{\boldsymbol{1}}_{n}) = C({\boldsymbol{\nu}}_{n},{\boldsymbol{1}}_{n})\int_{0}^{\infty}\dif\beta_{k}'\int_{0}^{\infty}\left(\prod_{j\neq k}{\dif\beta_{j}}\right)\int_{0}^{\infty}\left(\prod_{j=0}^{n}\alpha_{j}^{\nu_{j}-1}\,\dif\alpha_{j}\right)\,\mathcal{X}\delta(\mathcal{X}-\mathcal{Y})\\
    \quad\times\left\{ \left[\boldsymbol{1}^{T}\left(\boldsymbol{\alpha}+{\boldsymbol{\beta}}^{\left(k\right)}\right)-q\frac{\beta_{k}'}{q_{k}}\right]\alpha_{0}+\left(\boldsymbol{\alpha}+{\boldsymbol{\beta}}^{\left(k\right)}\right)^{T}\boldsymbol{Z}\left(\boldsymbol{\alpha}+{\boldsymbol{\beta}}^{\left(k\right)}\right)\right\} ^{\lambda_{0}}\,,
\end{multline}   
where
\begin{equation}
    \beta_{k}'\equiv\beta_{k}\left(1-W\,\frac{2\alpha_{l}+2\beta_{l}+\beta_{k}/q_{k}}{\alpha_{0}}\right) .
\end{equation}
Therefore
\begin{equation}\label{eq:h-expr}
    h_{k}({\boldsymbol{\nu}}_{n},{\boldsymbol{1}}_{n})=\frac{1}{\nu_{0}-1}\left(-\frac{q_{k}}{q}\right)\sum_{j=1}^{n}\left[\hat{j}^{+}\hat{k}^{-}I({\boldsymbol{\nu}}_{n}+{\boldsymbol{1}}_{n})\right]\,.
\end{equation}

Combining Eq.~\eqref{eq:g-expr}, \eqref{eq:g-I-h} and \eqref{eq:h-expr}, we finally obtain
\begin{equation}\label{eq:final_All}
    2\det(\boldsymbol{Z}) \, \hat{l}^{+}I(\boldsymbol{\nu}_{n}) = \sum_{j=1}^{n}\left[{Z_{j,l}}(\nu_{0}-\nu_{j})\right]I(\boldsymbol{\nu}_{n}) - \sum_{k=1}^{n} Z_{k,l}J_{k}({\boldsymbol{\nu}}_{n}) \,.
\end{equation}
where we have performed a simple shift of indices $\nu_i \to \nu_i-1$, and
\begin{equation}
    J_{k}(\boldsymbol{\nu}_{n})\equiv\sum_{j\ne k}\hat{j}^{+}\hat{k}^{-}I(\boldsymbol{\nu}_{n}) \,.
\end{equation}
In the context of integral reduction, we regard the integral which has a smaller total positive indices as ``simpler''. In this sense, it can be seen that the integrals appearing on the right-hand side of Eq.~\eqref{eq:final_All} are simpler than the one on the left-hand side. Since we are dealing with an irreducible sector, we have $\det(\bm{Z}) \neq 0$. Therefore, we can use Eq.~\eqref{eq:final_All} to recursively reduce a particular Feynman integral to simpler integrals.

\subsubsection{When the \texorpdfstring{$(l,l)$}{}-submatrix is singular}

In this case, the $(l,l)$-submatrix is non-invertible, which implies that $Z_{l,l}=0$. Eq.~\eqref{eq:q_define} then becomes singular. However, we find that the correct reduction rules can be obtained by taking the limit $Z_{l,l}\to0$ in the final expression Eq.~\eqref{eq:final_All}. Therefore, in this case, we have
\begin{equation}
    2\det(\boldsymbol{Z}) \, \hat{l}^{+}I(\boldsymbol{\nu}_{n})=\sum_{j\neq l}\left[{Z_{j,l}}(\nu_{0}-\nu_{j})\right]I(\boldsymbol{\nu}_{n})-\sum_{k\neq l}\left[Z_{k,l}J_{k}({\boldsymbol{\nu}}_{n})\right],
\end{equation}
which can be used to reduce $\hat{l}^{+}I(\boldsymbol{\nu}_{n})$.

A special case occurs when the matrix $\boldsymbol{Z}$ has only one element, meaning that $\boldsymbol{Z}(\hat{l},\hat{l})$ is an empty matrix. In this case, we have $\nu = \nu_1$ and $\nu_0 = d - \nu$. The procedure can stilled be carried out by defining $Z_{1,1} = 1$, and we obtain the correct reduction formula for single-propagator integrals:
\begin{equation}
    I(\nu+1)=\frac{d-2\nu}{2\nu \, \det(\boldsymbol{Z})}I(\nu)\,.
\end{equation}

\subsection{The reducible sectors}\label{sec:red-sector}

We now turn to the case where $\det(\bm{Z}) = 0\,$, which corresponds to reducible sectors according to the discussions in Sec.~\ref{sec:single}. Note that this criterion is equivalent to those found in earlier literatures \cite{Denner:2005nn}. The Gram matrix $\boldsymbol{Z}$ is singular in this case, and it implies that the degrees of freedom in the polynomial $\mathcal{F}$ are less than the number of variables. In another words, we can find a vector $\bm{\xi} = (\xi_1,\cdots,\xi_n)$ with $\xi_l = 1$, such that
\begin{equation}\label{eq:F-reducible}
    \mathcal{F}(\alpha_{1}+\xi_{1}\alpha_{l},\ldots,\alpha_{l},\ldots,\alpha_{n}+\xi_{n}\alpha_{l})=\mathcal{F}(\alpha_{1},\ldots,0,\ldots,\alpha_{n})\,.
\end{equation}
The vector $\bm{\xi}$ is in the kernel of $\bm{Z}$, i.e., $\bm{Z} \bm{\xi} = \bm{0}$.

\subsubsection{Reduction for general reducible sectors}\label{sec:redSec-All}

In Section~\ref{sec:single}, we have discussed a special case where the $\mathcal{F}$ polynomial already takes the form of the right-hand side of Eq.~\eqref{eq:F-reducible}, without introducing the transformation induced by $\bm{\xi}$. In other words, it corresponds to the case where all elements in $\bm{\xi}$, except $\xi_{l}=1$, are zero. In that case, integrals in this sector can be reduced to a single sub-sector.  More generically, we need to apply the variable change
\begin{equation}
    \alpha_{j}\to\alpha_{j}+\xi_{j}\alpha_{l}\,,\quad (\forall j\neq 0,l)\,,
\end{equation}
to transform the $\mathcal{F}$ polynomial into the desired form. Consequently, the integrals will be reduced to linear combinations of different sub-sectors, as we will demonstrate in the following.

After the variable change, an integral can be written as
\begin{align}\label{eq:I-change}
    I({\boldsymbol{\nu}}_{n}) & =C({\boldsymbol{\nu}}_{n})\int_{0}^{\infty}\left(\prod_{j=0}^{n}\alpha_{j}^{\nu_{j}-1}\,\dif\alpha_{j}\right)\,\mathcal{X}\delta(\mathcal{X}-\mathcal{Y})\left[\alpha_{0}\,\mathcal{U}+\mathcal{F}\right]^{\lambda_{0}}\nonumber\\
    &= C({\boldsymbol{\nu}}_{n})\int_{0}^{\infty}\alpha_{l}^{\nu_{l}-1}\,\dif\alpha_{l}\left(\prod_{j\neq0,l}\int_{-\xi_{j}\alpha_{l}}^{\infty}\left(\alpha_{j}+\xi_{j}\alpha_{l}\right)^{\nu_{j}-1}\,\dif\alpha_{j}\right)\nonumber\\
    & \quad\times\int_{0}^{\infty}\dif\alpha_{0}\,\alpha_{0}^{\nu_{0}-1}\left[\left(\mathcal{U}_{l,0}+\xi\alpha_{l}\right)\alpha_{0}+\mathcal{F}_{l,0}\right]^{\lambda_{0}}\,\mathcal{X}\delta(\mathcal{X}-\mathcal{Y}) \,, 
\end{align}
where $\xi\equiv\sum_{j=1}^{n}\xi_{j}$, $\mathcal{F}_{l,0} \equiv \mathcal{F}\big|_{\alpha_l \to 0}$ and we recall that $\mathcal{U}_{l,0} \equiv \mathcal{U}\big|_{\alpha_l \to 0}\,$. We can again split the integration domain using Eq.~\eqref{eq:integral_equal}, and perform a variable rescaling similar to that leading to Eq.~\eqref{eq:g-I-h}. We define
\begin{multline}
    G_{l}({\boldsymbol{\nu}}_{n})\equiv C({\boldsymbol{\nu}}_{n})\int_{0}^{\infty}\dif\alpha_{l}\,\alpha_{l}^{\nu_{l}-1}\int_{0}^{\infty}\dif\alpha_{0}\,\alpha_{0}^{\nu_{0}-1}\,\mathcal{X}\delta(\mathcal{X}-\mathcal{Y})\\
    \times\int_{0}^{\infty}\left(\prod_{j\neq0,l}\dif\alpha_{j} \left(\alpha_{j}+\xi_{j}\frac{\alpha_{l}}{\xi_{l}}\right)^{\nu_{j}-1}\right)\left[\left(\mathcal{U}_{l,0}+\xi\frac{\alpha_{l}}{\xi_{l}}\right)\alpha_{0}+\mathcal{F}_{l,0}\right]^{\lambda_{0}}\,,
\end{multline}
and
\begin{multline}
    H_{k}({\boldsymbol{\nu}}_{n})\equiv C({\boldsymbol{\nu}}_{n})\int_{0}^{\infty}\dif\alpha_{l}\,\alpha_{l}^{\nu_{l}-1}\int_{-\xi_{k}\alpha_{l}}^{0}{\dif\alpha_{k}} \left(\alpha_{k}+\xi_{k}\alpha_{l}\right)^{\nu_{k}-1}\int_{0}^{\infty}\dif\alpha_{0}\,\alpha_{0}^{\nu_{0}-1}\,\mathcal{X}\delta(\mathcal{X}-\mathcal{Y})\\
    \times\left(\prod_{j\neq0,l,k}\int_{\xi_{j}\alpha_{k}/\xi_{k}}^{\infty}{\dif\alpha_{j}} \left(\alpha_{j}+\xi_{j}\alpha_{l}\right)^{\nu_{j}-1}\right)\left[\left(\mathcal{U}_{l,0}+\xi\alpha_{l}\right)\alpha_{0}+\mathcal{F}_{l,0}\right]^{\lambda_{0}}\,.
\end{multline}
Let $S_{\xi}$ be the subset of $\{1,\cdots,n\}$ such that $\xi_k \neq 0$ if $k \in S_\xi$, we can then write
\begin{equation}
    I({\boldsymbol{\nu}}_{n}) = G_{l}({\boldsymbol{\nu}}_{n})+\sum_{k\in S_{\xi}/\{l\}}H_{k}({\boldsymbol{\nu}}_{n})\,.
\end{equation}

We can further show that $H_{k}({\boldsymbol{\nu}}_{n})$ can actually be rewritten as $G_{k}({\boldsymbol{\nu}}_{n})$. For that we introduce the variable change
\begin{align}
    &\alpha_{0}\to\alpha_{0} \, \frac{\alpha_{l}}{\alpha_{l}-\alpha_{k}/\xi_{k}} \,, \quad \alpha_{k}\to-\xi_{k}\alpha_{l} \, \frac{\alpha_{l}}{\alpha_{l}-\alpha_{k}/\xi_{k}} \,, \nonumber\\
    &\alpha_{j}\to\left(\alpha_{j}-\xi_{j}\alpha_{l}\right)\frac{\alpha_{l}}{\alpha_{l}-\alpha_{k}/\xi_{k}} \,, \quad (\forall j\neq 0,k,l) \,,
\end{align}
and therefore
\begin{multline}
    H_{k}({\boldsymbol{\nu}}_{n}) = C({\boldsymbol{\nu}}_{n})\int_{0}^{\infty}\dif\alpha_{k}\,\alpha_{k}^{\nu_{k}-1}\int_{0}^{\infty}\dif\alpha_{0}\,\alpha_{0}^{\nu_{0}-1}\,\mathcal{X}\delta(\mathcal{X}-\mathcal{Y})\\
    \times\int_{0}^{\infty}\left(\prod_{j\neq0,k}\dif\alpha_{j}\,\left(\alpha_{j}+\xi_{j}\frac{\alpha_{k}}{\xi_{k}}\right)^{\nu_{j}-1}\right)\left[\left(\mathcal{U}_{k,0}+\xi\frac{\alpha_{k}}{\xi_{k}}\right)\alpha_{0}+\mathcal{F}_{k,0}\right]^{\lambda_{0}}\,.
\end{multline}    
Using the method in Section~\ref{sec:single}, we can reduce $G_{k}({\boldsymbol{\nu}}_{n})$ (for both $k=l$ and $k\neq l$) to integrals in a sub-sector. Combining these results, we arrive at the reduction of $I({\boldsymbol{\nu}}_{n})$:
\begin{equation}\label{eq:redSec-final}
    I({\boldsymbol{\nu}}_{n})=\sum_{k \in S_{\xi}}\frac{\left(\xi_{k}\right)^{\nu_{k}}}{\left(\nu_{k}-1\right)!}\hat{k}_{0}\hat{P}_{k}I({\boldsymbol{\nu}}_{n})\,,
\end{equation}
where the operator $\hat{P}_{k}$ is a combination of index-raising and lowering operators:
\begin{equation}
    \begin{aligned}
        \hat{P}_{k}=\left(\prod_{j\in S_{\xi}}^{j\neq k}\sum_{\mu_{j}=0}^{\nu_{j}-1}\right)\left[B(-\nu_{0}+1,\mu+\nu_{k})\left(\frac{\sum_{i\neq k}\hat{i}^{+}}{-\xi}\right)^{\mu+\nu_{k}} \prod_{m\in S_{\xi}}^{m\neq k}\frac{\left(\xi_{m}\hat{m}^{-}\right)^{\mu_{m}}}{\mu_{m}!} \right]\,,
    \end{aligned}
\end{equation}
where $\mu\equiv\sum_{j\in S_{\xi}}^{j\neq k}\mu_{j}$. Obviously, the above formula only applies if $\xi \neq 0$. The case for $\xi = 0$ will be discussed in the following.

\subsubsection{Degenerate limits}

The case $\xi = 0$ corresponds to certain degenerate limits of Feynman integrals. As a specific example, consider the bubble diagram shown in Figure~\ref{fig:bubble} with $m_{1}^{2}=m_{2}^{2}$ and $s=0$. Here we find that $\bm{\xi}=\{1,-1\}$, and hence $\xi=0$. We now discuss how to deal with this kind of situations.

When $\xi=0\,$, both $\mathcal{U}$ and $\mathcal{F}$ satisfy
\begin{equation}
    \mathcal{O}(\alpha_{1},\cdots,\alpha_{l},\cdots,\alpha_{n})  =\mathcal{O}(\alpha_{1}-\xi_{1}\alpha_{l},\cdots,0,\cdots,\alpha_{n}-\xi_{n}\alpha_{l})\,,
\end{equation}
for certain $l$, where $\mathcal{O} \in \{\mathcal{U},\mathcal{F}\}$. Therefore $\alpha_0 \, \mathcal{U} + \mathcal{F}$ satisfies the above property as well. We now consider the integral $I(\bm{\nu}_n+\bm{1}_n)$. By introducing the $\beta$-variables as in Section~\ref{sec:Decrease}, we can write it as
\begin{equation}
    I({\boldsymbol{\nu}}_{n},{\boldsymbol{1}}_{n}) = C({\boldsymbol{\nu}}_{n},{\boldsymbol{1}}_{n})\int_{0}^{\infty}\left(\prod_{j=1}^{n}{\dif\beta_{j}}\right)\int_{0}^{\infty}\left(\prod_{j=0}^{n}\alpha_{j}^{\nu_{j}-1}\,\dif\alpha_{j}\right){\mathcal{G}_{0}(\boldsymbol{\alpha},\boldsymbol{\beta})}^{\lambda_{0}}\,\mathcal{X}\delta(\mathcal{X}-\mathcal{Y})\,,
\end{equation} 
where
\begin{equation}
    \mathcal{G}_{0}(\boldsymbol{\alpha},\boldsymbol{\beta}) \equiv \left[ \alpha_0 \, \mathcal{U} + \mathcal{F} \right]_{\bm{\alpha} \to \bm{\alpha}+\bm{\beta}} = \left[\boldsymbol{1}^{T}\left(\boldsymbol{\alpha}+\boldsymbol{\beta}\right)\right]\alpha_{0}+{\left(\boldsymbol{\alpha}+\boldsymbol{\beta}\right)}^{T}\boldsymbol{Z}\left(\boldsymbol{\alpha}+\boldsymbol{\beta}\right) .
\end{equation}
It follows from the property of $\mathcal{U}$ and $\mathcal{F}$ that
\begin{equation}
    \mathcal{G}_{0}(\boldsymbol{\alpha},\beta_{1},\cdots,\beta_{l},\cdots,\beta_{n})=\mathcal{G}_{0}({\boldsymbol{\alpha}},\beta_{1}-\xi_{1}\beta_{l},\cdots,0,\cdots,\beta_{n}-\xi_{n}\beta_{l})\,.
\end{equation}
We now show that this property can be used for reducing the integral to sub-sectors.

For convenience, we consider a function $g(y,\bm{x}_{n-1})$ which satisfies
\begin{equation}
    g(y,\bm{x}_{n-1})=g(0,\bm{x}_{n-1}+y\bm{\eta}_{n-1}) \,,
\end{equation}
where $\bm{x}_{n-1}$ denotes the sequence of $n-1$ variables, and $\bm{\eta}_{n-1}$ is the sequence of $n-1$ constants. Using the relation
\begin{equation}
    \int_{0}^{\infty}\dif x_{l}\int_{x_{l}}^{\infty}\left(\prod_{j\neq l}\dif x_{j}\right)=\sum_{k\neq l}\int_{0}^{\infty}\dif x_{k}\int_{x_{k}}^{\infty}\left(\prod_{j\neq k,l}\dif x_{j}\right)\int_{0}^{x_{k}}\dif x_{l}\,,
\end{equation}
we can show that
\begin{align}\label{eq:g_decomp}
    \int_{0}^{\infty}\dif y\int_{0}^{\infty}\left(\prod_{j=1}^{n-1}\dif x_{j}\right)g(y,\bm{x}_{n-1}) & =\int_{0}^{\infty}\dif y\left(\prod_{j=1}^{n-1}\int_{y\eta_{j}}^{\infty}\dif x_{j}'\right)g(0,\bm{x}_{n-1}')\nonumber\\
    & =\sum_{k=1}^{n-1}\int_{0}^{\infty}\dif x_{k}'\left(\prod_{j=1}^{n-1}\int_{\eta_{j}x_{k}'/\eta_{k}}^{\infty}\dif x_{j}'\right)\int_{0}^{x_{k}'}\dif y\,g(0,\bm{x}_{n-1}')\nonumber\\
    & =\sum_{k=1}^{n-1}\eta_{k}\int_{0}^{\infty}\dif Y\int_{0}^{\infty}\left(\prod_{j\neq k}\dif X_{j}\right)Y\,g(Y,\bm{X}_{n-1})\,,
\end{align}
where $\bm{x}_{n-1}' \equiv \bm{x}_{n-1}+y \bm{\eta}_{n-1}$, $Y\equiv x_{k}'/\eta_{k} = x_{k}/\eta_{k}+y$, and $X_{j}\equiv x_{j}'-\eta_{j} x_{k}'/\eta_{k} = x_{j}-\eta_{j} x_{k}/\eta_{k}$. Note that in the $k$th term of the above result, we have $X_k = 0$ in $g$, and we don't have to integrate over $X_k$. Hence, we have eliminated one variable at the cost of increasing the power of $Y$.

Applying Eq.~\eqref{eq:g_decomp} to $I({\boldsymbol{\nu}}_{n},{\boldsymbol{1}}_{n})$, we get 
\begin{multline}
    I({\boldsymbol{\nu}}_{n},{\boldsymbol{1}}_{n})=-\sum_{k\neq l}\xi_{k}C({\boldsymbol{\nu}}_{n},{\boldsymbol{1}}_{n})\int_{0}^{\infty}\left(\prod_{j\neq k}{\dif\beta_{j}}\right)\int_{0}^{\infty}\left(\prod_{j=0}^{n}\alpha_{j}^{\nu_{j}-1}\,\dif\alpha_{j}\right)\\
    \times\beta_{l}\,{\mathcal{G}_{0}(\boldsymbol{\alpha},{\boldsymbol{\beta}}^{\left(k\right)})}^{\lambda_{0}}\,\mathcal{X}\delta(\mathcal{X}-\mathcal{Y})\,.
\end{multline}
Absorbing the extra $\beta_{j}$ variables using Eq.~\eqref{eq:de_power_2}, we finally obtain the reduction formula that works in the degenerate limits:
\begin{equation}\label{eq:degenerate_lim}
    I(\cdots,\nu_{l},\cdots)=-\sum_{k\neq l}\xi_{k}I(\cdots,\nu_{k}-1,\cdots,\nu_{l}+1,\cdots)\,.
\end{equation}
It can be observed that by fixing $l$, we can recursively decrease other indices at the cost of increasing $\nu_l$. This eventually leads to integrals in the sub-sectors.

Now, let us consider the example mentioned earlier: the bubble diagram with $m_{1}^{2}=m_{2}^{2}$ and $s=0$. By choosing $l=1$ and repeatedly applying Eq.~\eqref{eq:degenerate_lim}, we can arrive at
\begin{equation}
    I(\nu_{1},\nu_{2})=I(\nu_{1}+\nu_{2},0)\,.
\end{equation}

Up to now, we have exhausted all possibilities and have presented iterative formulas to reduce any one-loop integral with indices larger than one to linear combinations of simpler integrals. The remaining task is to reduce the negative indices corresponding to numerators in the momentum representation Eq.~\eqref{eq:FeyInt}.

\subsection{Reduction of integrals with numerators}

If any index $\nu_j < 0$, the corresponding $D_j$ is in the numerator of Eq.~\eqref{eq:FeyInt}. Such integrals can always be reduced to integrals with $\nu_j = 0$. In the following, we show how to achieve this reduction within our approach.

\subsubsection{Rewriting the integrals in shifted dimensions}

It is well-known that integrals with $\nu_j < 0$ can be rewritten as integrals with $\nu_j = 0$ in shifted spacetime dimensions. This can be easily seen in the LP representation: 
\begin{equation}
    I^{(d)}({\boldsymbol{\nu}}_{n})=C^{(d)}({\boldsymbol{\nu}}_{n})\int_{0}^{\infty} \mathcal{G}^{-d/2} \prod_{j=1}^{n} \alpha_{j}^{\nu_{j}-1} {\dif\alpha_{j}} \,,
\end{equation}
where $\mathcal{G}\equiv\mathcal{U}+\mathcal{F}$. We have used the superscript $(d)$ to specify the dependence on the spacetime dimension.

If $\nu_j < 0$, we can increase $\nu_j$ with a simple integration-by-parts:
\begin{equation}
    \int_{0}^{\infty}\dif\alpha_{j}\frac{\alpha_{j}^{\rho+\nu_{j}-1}}{\Gamma(\rho+\nu_{j})}\mathcal{G}^{-d/2} = \frac{d}{2}\int_{0}^{\infty}\dif\alpha_{j}\frac{\alpha_{r}^{\rho+\nu_{j}}}{\Gamma(\rho+\nu_{j}+1)}\left(\frac{\partial\mathcal{G}}{\partial\alpha_{j}}\right)\mathcal{G}^{-(d+2)/2}\,,
\end{equation}    
where we have explicitly introduced the regulator $\rho$ as discussed in Section~\ref{sec:NewCW}. One may see that the right-hand side of the above equation corresponds to integrals in $d+2$ dimensions, but with an increased $\nu_j$ index. Note that $\mathcal{U}$ and $\mathcal{F}$ are homogeneous polynomials of degree $L$ and $L+1$, respectively. The above relation can be rewritten in terms of index-raising operators as
\begin{equation}\label{eq:numerator}
    I^{(d)}(\cdots,\nu_{j},\cdots)=(-1)^{L}\left[\left((L+1)\frac{d}{2}-\sum_{i=1}^{n}\nu_{i}\right)\,\hat{\mathcal{U}}_{j}^{+}-\hat{\mathcal{F}}_{j}^{+}\right]I^{(d+2)}(\cdots,\nu_{j}+1,\cdots)\,,
\end{equation}
where $\hat{\mathcal{O}}_{j}^{+}$ (for $\mathcal{O} \in \{\mathcal{U},\mathcal{F}\}$) represents an operator obtained by substituting each $\alpha_{k}$ in $\left(\partial\mathcal{O}/\partial\alpha_{j}\right)$ by $\hat{k}^{+}$. Note that Eq.~\eqref{eq:numerator} holds for general $L$-loop integrals, and for the purpose of this work, we can set $L = 1\,$.

With Eq.~\eqref{eq:numerator}, we can recursively increase a negative index to zero, at the cost of employing integrals in shifted dimensions. Let us consider a simple example: $I(-1,2)$ from the bubble integral family in Figure~\ref{fig:bubble}. Applying Eq.~\eqref{eq:numerator} with $j=1$, we obtain 
\begin{equation}\label{eq:redf_12}
    I^{(d)}(-1,2)=\left(1-d\right)I^{(d+2)}(0,2)+2\left(m_{1}^{2}+m_{2}^{2}-s\right)I^{(d+2)}(0,3)\,.
\end{equation}

\subsubsection{Dimensional recurrence relations}

We now need to transform the integrals in shifted dimensions to the $d$-dimensional ones. This is the so-called dimensional recurrence relations~\cite{Tarasov:1996br, Tarasov:1997kx, Lee:2009dh, Lee:2010ug}. We show how such relations can be naturally derived within our approach.

In the Feynman parameterization, it is easy to increase the dimension:
\begin{equation}
    I^{(d)}({\boldsymbol{\nu}}_{n})=-\sum_{k=1}^{n}\hat{k}^{+}I^{(d+2)}({\boldsymbol{\nu}}_{n})\,.
\end{equation}
We need to find the inverse relation, to express $I^{(d+2)}$ in terms of $I^{(d)}$. Note that the indices here are all non-negative. Since we can apply the reduction rules that we have constructed in both $d$ and $d+2$ dimensions, we only need to find the relations among corner integrals. Combining the above equation with Eq.~\eqref{eq:final_All}, we obtain 
\begin{align}\label{eq:dim-shift}
    2\det(\boldsymbol{Z})I^{(d)}(\boldsymbol{1}_{n}) &= -2\det(\boldsymbol{Z})\sum_{l=1}^{n}\hat{l}^{+}I^{(d+2)}(\boldsymbol{1}_{n})\nonumber
    \\
    &=  \zeta \left(n-1-d\right)I^{(d+2)}(\boldsymbol{1}_{n})-\sum_{k=1}^{n} \zeta_{k}\hat{k}_{0}I^{(d)}(\boldsymbol{1}_{n})\,,
\end{align}
where $\zeta_{k}\equiv\sum_{j=1}^{n}Z_{k,j}$, $\zeta\equiv\sum_{k=1}^{n}\zeta_{k}$, and we have used
\begin{equation}
    \sum_{j\ne k} \hat{j}^{+}\hat{k}_{0}I^{(d+2)}(\boldsymbol{1}_{n}) =-\hat{k}_{0}I^{(d)}(\boldsymbol{1}_{n})\,.
\end{equation}
If $\zeta\neq0$, Eq.~\eqref{eq:dim-shift} gives rise to the dimension recurrence relation that we will need:
\begin{equation}\label{eq:dim-shift-z!=0}
    I^{(d+2)}(\boldsymbol{1}_{n})=\frac{2\det(\boldsymbol{Z})}{\zeta\left(n-1-d\right)}I^{(d)}(\boldsymbol{1}_{n})+\frac{1}{\zeta\left(n-1-d\right)}\sum_{k=1}^{n}\zeta_{k}\hat{k}_{0}I^{(d)}(\boldsymbol{1}_{n})\,.
\end{equation}
Combining with Eq.~\eqref{eq:numerator}, we complete the reduction of integrals with numerators. 

Applying the above relation, we can continue the reduction of $I(-1,2)$ starting from Eq.~\eqref{eq:redf_12}. We first reduce the right-hand side of Eq.~\eqref{eq:redf_12} to corner integrals in $d+2$ dimensions, and obtain 
\begin{equation}
    I^{(d)}(-1,2)=-\frac{d}{2m_{2}^{2}}\frac{2m_{2}^{2}-\left(d-2\right)\left(m_{1}^{2}-m_{2}^{2}-s\right)}{2m_{2}^{2}}I^{(d+2)}(0,1)\,.
\end{equation}
We then use Eq.~\eqref{eq:dim-shift-z!=0} to derive (recall that we have defined $Z_{1,1}=1$ when the matrix $\bm{Z}$ has only one element)
\begin{equation}
    I^{(d+2)}(0,1)=-\frac{2 m_{2}^{2}}{d}I^{(d)}(0,1)\,.
\end{equation}
Finally, we have 
\begin{equation}
    I^{(d)}(-1,2)=\frac{2m_{2}^{2}-\left(d-2\right)\left(m_{1}^{2}-m_{2}^{2}-s\right)}{2m_{2}^{2}}I^{(d)}(0,1)\,.
\end{equation}

Note that if $\zeta = 0$, Eq.~\eqref{eq:dim-shift} does not lead to a dimension recurrence, but gives rise to an interesting reduction rule of the corner integral to sub-sectors:
\begin{equation}\label{eq:magic}
    I^{(d)}(\boldsymbol{1}_{n})=-\frac{1}{2\det(\boldsymbol{Z})}\sum_{k=1}^{n}\zeta_{k}\hat{k}_{0}I^{(d)}(\boldsymbol{1}_{n})\,.
\end{equation}
This is the so-called ``magic relation''. In the usual IBP reduction, it can only be observed when IBP relations involving certain super-sectors are included~\cite{Klappert:2020nbg, Frellesvig:2018ymi}. 
For example, for the bubble diagram in Figure~\ref{fig:bubble}, we find $\zeta=s$. Therefore, when $s=0$, the integral $I(1,1)$ can be reduced to sub-sectors:
\begin{equation}
    I^{(d)}(1,1)=\frac{1}{m_{1}^{2}-m_{2}^{2}}\left[I^{(d)}(1,0)-I^{(d)}(0,1)\right]\,.
\end{equation}
In \texttt{Kira}, this reduction rule can only be found when embedding the bubble integrals in a triangle super-sector.

\section{Examples}\label{sec:example}

\subsection{One-loop integrals}

The recursive formulas derived in the previous Section allow a straightforward computer implementation. We have written a proof-of-concept \texttt{Mathematica} code to automatically reduce one-loop integrals. We have tested the program with various examples and confirmed its correctness.

\begin{figure}[!th]
	\centering
	\includegraphics[width=0.41\linewidth]{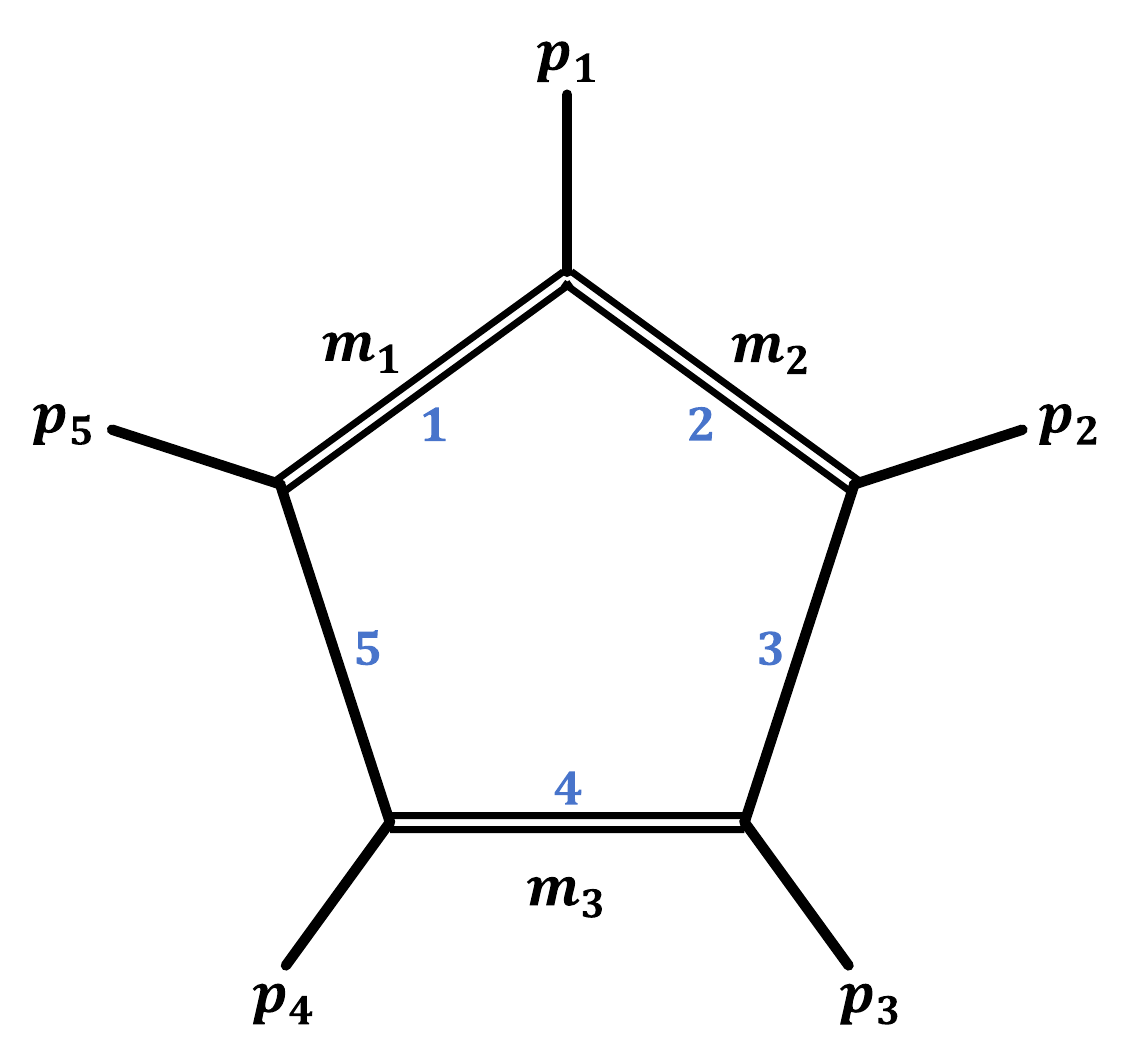}
	\caption{The pentagon diagram with three masses.}
    \label{fig:pentagon}
\end{figure}

As a more complicated example, consider the pentagon integral family with three masses shown in Figure~\ref{fig:pentagon}. We set all external momenta to be light-like. The kinematic variables are $m_{1}^{2}$, $m_{2}^{2}$, $m_{3}^{2}$, $(p_{1}+p_{2})^2=s_{12}$, $(p_{2}+p_{3})^2=s_{23}$, $(p_{3}+p_{4})^2=s_{34}$, $(p_{4}+p_{5})^2=s_{45}$ and $(p_{5}+p_{1})^2=s_{51}$.
The matrix $\boldsymbol{Z}$ in this family is 
\begin{equation}
    \begin{pmatrix}
        m_{1}^{2} & \frac{1}{2}\left(m_{1}^{2}+m_{2}^{2}\right) & \frac{1}{2}\left(m_{1}^{2}-s_{12}\right) & \frac{1}{2}\left(m_{1}^{2}+m_{3}^{2}-s_{45}\right) & \frac{1}{2}m_{1}^{2}\\
        \frac{1}{2}\left(m_{1}^{2}+m_{2}^{2}\right) & m_{2}^{2} & \frac{1}{2}m_{2}^{2} & \frac{1}{2}\left(m_{2}^{2}+m_{3}^{2}-s_{23}\right) & \frac{1}{2}\left(m_{2}^{2}-s_{51}\right)\\
        \frac{1}{2}\left(m_{1}^{2}-s_{12}\right) & \frac{1}{2}m_{2}^{2} & 0 & \frac{1}{2}m_{3}^{2} & -\frac{1}{2}s_{34}\\
        \frac{1}{2}\left(m_{1}^{2}+m_{3}^{2}-s_{45}\right) & \frac{1}{2}\left(m_{2}^{2}+m_{3}^{2}-s_{23}\right) & \frac{1}{2}m_{3}^{2} & m_{3}^{2} & \frac{1}{2}m_{3}^{2}\\
        \frac{1}{2}m_{1}^{2} & \frac{1}{2}\left(m_{2}^{2}-s_{51}\right) & -\frac{1}{2}s_{34} & \frac{1}{2}m_{3}^{2} & 0
    \end{pmatrix}
    .
\end{equation}

We consider the reduction of $I(1,2,1,1,1)$. Since $\boldsymbol{Z}$ is non-singular, we should apply Eq.~\eqref{eq:final_All} with $\boldsymbol{\nu}_{n}=(1,1,1,1,1)$ and $l=2$, which leads to 
\begin{equation}
    I(1,2,1,1,1)=\left(d-6\right)\frac{\det(\boldsymbol{M})}{2\det(\boldsymbol{Z})}I(1,1,1,1,1)+\text{sub-sector integrals}\,,
\end{equation}
where $\boldsymbol{M}$ represents the matrix formed by replacing all elements in the second row of $\boldsymbol{Z}$ with $1$. The sub-sector integrals in the above expression may still contain integrals that need to be reduced (e.g., $I(2,1,1,1,0)$, $I(0,1,2,1,1)$, etc.). We can apply the relevant reduction formulas recursively until only master integrals remain. We have verified that the results agree with those provided by \texttt{Kira}.

We can also consider a reducible case, with $m_{1}^{2}=m_{2}^{2}=m_{3}^{2}=s_{12}=0$. At this kinematic point, the $\boldsymbol{Z}$ matrix is singular. Therefore, we need to apply Eq.~\eqref{eq:redSec-final} when performing the reduction of $I(1,2,1,1,1)$. The $\bm{\xi}$ vector can be derived as
\begin{equation}
    \boldsymbol{\xi}=\left\{-\frac{s_{23}}{s_{45}},1,-\frac{s_{51}}{s_{34}},0,0\right\} .
\end{equation}
It follows that $I(1,2,1,1,1)$ can be reduced to integrals in 3 sub-sectors:
\begin{align}
    I(1,2,1,1,1) & =\frac{1}{(d-7)(d-8)}\frac{2s_{34}^{2}s_{45}^{2}}{(s_{23}s_{34}-s_{34}s_{45}+s_{45}s_{51})^{2}} \, \big[ I(2,0,2,1,1)+I(2,0,1,2,1) \nonumber\\
    & \quad +I(2,0,1,1,2)+I(1,0,2,2,1)+I(1,0,2,1,2)+I(1,0,1,2,2) \nonumber\\
    & \quad +I(3,0,1,1,1)+I(1,0,3,1,1)+I(1,0,1,3,1)+I(1,0,1,1,3) \big] \nonumber\\
    & \quad+\text{integrals in the other two sub-sectors} \,,
\end{align}
where the other two sub-sectors are those where the first or the third propagator is removed (e.g., $I(0,1,1,2,2)$, $I(1,1,0,2,2)$, etc.).

Finally, we consider a degenerate limit, where $p_{1}=p_{2}$ in addition to $m_{1}^{2}=m_{2}^{2}=m_{3}^{2}=0$. In this case, we have $s_{12}=0$, $s_{23}=s_{45}/2$ and $s_{51}=s_{34}/2$. Clearly, the top-sector is reducible, but we cannot apply Eq.~\eqref{eq:redSec-final} because $\bm{\xi}=\{1,-2,1,0,0\}$, which implies $\xi=0$. Therefore, in this case, we need to apply Eq.~\eqref{eq:degenerate_lim}. By choosing $l=1$, we ultimately obtain 
\begin{equation}
    I(1,2,1,1,1)=4I(3,0,1,1,1)-2I(3,1,0,1,1)-I(2,2,0,1,1)\,.
\end{equation}

\subsection{A two-loop example}

\begin{figure}[th!]
    \centering
    \includegraphics[width=0.41\linewidth]{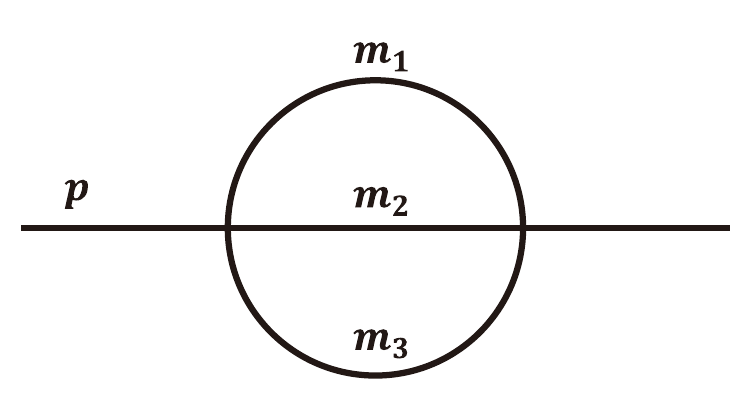}
    \caption{The sunrise diagram with $p^2 = s$.}
    \label{fig:sunrise}
\end{figure}

Given the success in one-loop problems, it is natural to extend our approach to multi-loop integrals. A new ingredient at higher loops is that there can be more than one MIs within a sector, and it is no longer guaranteed that all indices can be reduced to $1$ or $0$. Correspondingly, in our approach, we find that when we perform the variable changes, certain variables may appear in the denominators of the transformed integrands. Such transformations are therefore only valid if the corresponding indices are greater than $1$. In addition, the structure of degenerate limits is also much more complicated than the one-loop case.

As a simple example, consider the massless sunrise family shown in Figure~\ref{fig:sunrise}, with $m_1 = m_2 = m_3 = 0$. The integrals in this family can be represented as 
\begin{multline}
    I(\nu_{1},\nu_{2},\nu_{3})=C(\nu_{1},\nu_{2},\nu_{3})\int_{0}^{\infty}\dif\alpha_{0}\dif\alpha_{1}\dif\alpha_{2}\dif\alpha_{3}\,\alpha_{0}^{\nu_{0}-1}\alpha_{1}^{\nu_{1}-1}\alpha_{2}^{\nu_{2}-1}\alpha_{3}^{\nu_{3}-1}\mathcal{X}\delta(\mathcal{X}-\mathcal{Y})
    \\
    \times\left[(\alpha_{1}\alpha_{2}+\alpha_{1}\alpha_{3}+\alpha_{2}\alpha_{3})\alpha_{0}-\alpha_{1}\alpha_{2}\alpha_{3}s\right]^{\lambda_{0}}\,.
\end{multline}
Following the method in Section~\ref{sec:red-irred-sectors}, we define the auxiliary integral 
\begin{multline}
    g(\nu_{1},\nu_{2},\nu_{3})=C(\nu_{1},\nu_{2},\nu_{3})\int_{0}^{\infty}\dif\alpha_{0}\dif\alpha_{1}\dif\alpha_{2}\dif\alpha_{3}\,\alpha_{0}^{\nu_{0}-1}\alpha_{1}^{\nu_{1}-1}\alpha_{2}^{\nu_{2}-1}\alpha_{3}^{\nu_{3}-1}\mathcal{X}\delta(\mathcal{X}-\mathcal{Y})
    \\
    \times\left[(\alpha_{1}\alpha_{3}+\alpha_{2}\alpha_{3})\alpha_{0}\right]^{\lambda_{0}}\,,
    \label{eq:aux_two_loop}
\end{multline}
which is itself a scaleless integral and should therefore vanish. By performing a variable change $\alpha_1\to\alpha_1 (1+\alpha_2/\alpha_3-s\alpha_2/\alpha_0)\,$, we can express $g(\nu_{1},\nu_{2},\nu_{3})$ in terms of integrals in the top sector, thus establishing a reduction relation for top-sector integrals. Unlike the case in one-loop integrals, the variable change now introduces the variable $\alpha_3$ in the denominator. As a result, the auxiliary integral must have $\nu_3 \geq 2$ in order to yield a valid reduction relation. For instance, from the auxiliary integral $g(1,1,2)$, we obtain the correct reduction relation
\begin{equation}
    I(1,2,2)=\frac{10-3 d}{2 s}\left[I(1,2,1)+I(1,1,2)\right] .
\end{equation}

From IBP reduction, we know that $I(1,2,1)$ and $I(1,1,2)$ can be further reduced to $I(1,1,1)$. Such relations cannot be obtained through the auxiliary integral \eqref{eq:aux_two_loop}. In fact, this is one of the degenerate limits, which deserves a systematic investigation in the future. Furthermore, in the above example we have only applied variable transformations similar to the one-loop method. There are still significant flexibility in the integration contours yet to be fully explored.

\section{Summary and outlook}\label{sec:summary}

In this paper, we have explored interesting properties of the Feynman parameterization that are useful for Feynman integral reduction without employing IBP relations. In particular, we have derived an extension of the Cheng-Wu theorem, such that the delta-function in the Feynman parameterization can take a more general form. This can be regarded as a specific implementation of the equivalence of integration contours, which is related to the twisted homology of Feynman integrals. Leveraging these properties, we have derived universal recursive formulas for the reduction of one-loop integrals in both irreducible and reducible sectors, and our method works in degenerate limits as well. The reduction relations can be easily implemented in a computer program. We have applied them to various examples and observed remarkable performance.

We have also demonstrated the validity of our method at two loops. In an explicit two-loop example, the straightforward application of the one-loop method can yield the correct reduction relations. However, we also emphasize that not all reduction relations can be obtained in this way, since additional structures arise at higher loops. This warrants further investigation, including more ways to perform variable changes and contour deformations. In particular, our manipulations strongly depends on the structure of the matrix $\boldsymbol{Z}$, which is a Gram matrix of external momenta with entries being kinematic invariants. Its determinant, $\det(\boldsymbol{Z})$, corresponds to a Landau singularity~\cite{Landau:1959fi} and appears in the denominator of the reduction coefficients. The relation between Landau singularities and reduction coefficients has been studied in, e.g., \cite{Coro:2025kha}. These singularities can be obtained systematically with dedicated packages such as \texttt{Baikovletter}~\cite{Jiang:2024eaj} or \texttt{SOFIA}~\cite{Correia:2025wtb}. It remains to be investigated whether the knowledge of the Landau singularities can help us to design variable transformations for integral reduction.

Furthermore, in this work we have only exploited the equivalence of integration contours in a simple way. There can be deeper mathematical structures behind these equivalence relations. For example, it is possible to employ the concept of intersection numbers between two integration contours to directly compute the reduction coefficients. This provides a particularly intriguing future perspective.

\begin{acknowledgments}
The authors would like to thank Liangliang Zhao, Tingfei Li, Mingming Lu and Yiyang Zhang for useful discussions. This work was supported in part by the National Natural Science Foundation of China under Grant No. 12375097, 12535003, 12547104, and the Fundamental Research Funds for the Central Universities.
\end{acknowledgments}

\bibliographystyle{JHEP}
\bibliography{references_inspire.bib,references_local.bib}

\providecommand{\href}[2]{#2}\begingroup\raggedright\begin{thebibliography}{10}

\bibitem{Kotikov:1990kg}
A.V.~Kotikov, \emph{{Differential equations method: New technique for massive
  Feynman diagrams calculation}},
  \href{https://doi.org/10.1016/0370-2693(91)90413-K}{\emph{Phys. Lett. B}
  {\bfseries 254} (1991) 158}.

\bibitem{Kotikov:1991pm}
A.V.~Kotikov, \emph{{Differential equation method: The Calculation of N point
  Feynman diagrams}},
  \href{https://doi.org/10.1016/0370-2693(91)90536-Y}{\emph{Phys. Lett. B}
  {\bfseries 267} (1991) 123}.

\bibitem{Remiddi:1997ny}
E.~Remiddi, \emph{{Differential equations for Feynman graph amplitudes}},
  \href{https://doi.org/10.1007/BF03185566}{\emph{Nuovo Cim. A} {\bfseries 110}
  (1997) 1435} [\href{https://arxiv.org/abs/hep-th/9711188}{{\ttfamily
  hep-th/9711188}}].

\bibitem{Gehrmann:1999as}
T.~Gehrmann and E.~Remiddi, \emph{{Differential equations for two-loop
  four-point functions}},
  \href{https://doi.org/10.1016/S0550-3213(00)00223-6}{\emph{Nucl. Phys. B}
  {\bfseries 580} (2000) 485}
  [\href{https://arxiv.org/abs/hep-ph/9912329}{{\ttfamily hep-ph/9912329}}].

\bibitem{Laporta:2000dc}
S.~Laporta, \emph{{Calculation of master integrals by difference equations}},
  \href{https://doi.org/10.1016/S0370-2693(01)00256-8}{\emph{Phys. Lett. B}
  {\bfseries 504} (2001) 188}
  [\href{https://arxiv.org/abs/hep-ph/0102032}{{\ttfamily hep-ph/0102032}}].

\bibitem{Tkachov:1981wb}
F.V.~Tkachov, \emph{{A theorem on analytical calculability of 4-loop
  renormalization group functions}},
  \href{https://doi.org/10.1016/0370-2693(81)90288-4}{\emph{Phys. Lett. B}
  {\bfseries 100} (1981) 65}.

\bibitem{Chetyrkin:1981qh}
K.G.~Chetyrkin and F.V.~Tkachov, \emph{{Integration by parts: The algorithm to
  calculate $\beta$-functions in 4 loops}},
  \href{https://doi.org/10.1016/0550-3213(81)90199-1}{\emph{Nucl. Phys. B}
  {\bfseries 192} (1981) 159}.

\bibitem{Smirnov:2023yhb}
A.V.~Smirnov and M.~Zeng, \emph{{FIRE 6.5: Feynman integral reduction with new
  simplification library}},
  \href{https://doi.org/10.1016/j.cpc.2024.109261}{\emph{Comput. Phys. Commun.}
  {\bfseries 302} (2024) 109261}
  [\href{https://arxiv.org/abs/2311.02370}{{\ttfamily 2311.02370}}].

\bibitem{Lee:2013mka}
R.N.~Lee, \emph{{LiteRed 1.4: a powerful tool for reduction of multiloop
  integrals}}, \href{https://doi.org/10.1088/1742-6596/523/1/012059}{\emph{J.
  Phys. Conf. Ser.} {\bfseries 523} (2014) 012059}
  [\href{https://arxiv.org/abs/1310.1145}{{\ttfamily 1310.1145}}].

\bibitem{vonManteuffel:2012np}
A.~von Manteuffel and C.~Studerus, \emph{{Reduze 2 - Distributed Feynman
  Integral Reduction}},  \href{https://arxiv.org/abs/1201.4330}{{\ttfamily
  1201.4330}}.

\bibitem{Klappert:2020nbg}
J.~Klappert, F.~Lange, P.~Maierh\"ofer and J.~Usovitsch, \emph{{Integral
  reduction with Kira 2.0 and finite field methods}},
  \href{https://doi.org/10.1016/j.cpc.2021.108024}{\emph{Comput. Phys. Commun.}
  {\bfseries 266} (2021) 108024}
  [\href{https://arxiv.org/abs/2008.06494}{{\ttfamily 2008.06494}}].

\bibitem{Lee:2014tja}
R.N.~Lee, \emph{{Modern techniques of multiloop calculations}},  in \emph{{49th
  Rencontres de Moriond on QCD and High Energy Interactions}}, pp.~297--300,
  2014 [\href{https://arxiv.org/abs/1405.5616}{{\ttfamily 1405.5616}}].

\bibitem{Larsen:2015ped}
K.J.~Larsen and Y.~Zhang, \emph{{Integration-by-parts reductions from unitarity
  cuts and algebraic geometry}},
  \href{https://doi.org/10.1103/PhysRevD.93.041701}{\emph{Phys. Rev. D}
  {\bfseries 93} (2016) 041701}
  [\href{https://arxiv.org/abs/1511.01071}{{\ttfamily 1511.01071}}].

\bibitem{Bendle:2019csk}
D.~Bendle, J.~B\"ohm, W.~Decker, A.~Georgoudis, F.-J.~Pfreundt, M.~Rahn et~al.,
  \emph{{Integration-by-parts reductions of Feynman integrals using Singular
  and GPI-Space}}, \href{https://doi.org/10.1007/JHEP02(2020)079}{\emph{JHEP}
  {\bfseries 02} (2020) 079}
  [\href{https://arxiv.org/abs/1908.04301}{{\ttfamily 1908.04301}}].

\bibitem{Chen:2022jux}
J.~Chen and B.~Feng, \emph{{Module intersection and uniform formula for
  iterative reduction of one-loop integrals}},
  \href{https://doi.org/10.1007/JHEP02(2023)178}{\emph{JHEP} {\bfseries 02}
  (2023) 178} [\href{https://arxiv.org/abs/2207.03767}{{\ttfamily
  2207.03767}}].

\bibitem{Chen:2019mqc}
W.~Chen, \emph{{Reduction of Feynman Integrals in the Parametric
  Representation}}, \href{https://doi.org/10.1007/JHEP02(2020)115}{\emph{JHEP}
  {\bfseries 02} (2020) 115}
  [\href{https://arxiv.org/abs/1902.10387}{{\ttfamily 1902.10387}}].

\bibitem{Chen:2019fzm}
W.~Chen, \emph{{Reduction of Feynman Integrals in the Parametric Representation
  II: Reduction of Tensor Integrals}},
  \href{https://doi.org/10.1140/epjc/s10052-021-09036-5}{\emph{Eur. Phys. J. C}
  {\bfseries 81} (2021) 244}
  [\href{https://arxiv.org/abs/1912.08606}{{\ttfamily 1912.08606}}].

\bibitem{Chen:2020wsh}
W.~Chen, \emph{{Reduction of Feynman integrals in the parametric representation
  III: integrals with cuts}},
  \href{https://doi.org/10.1140/epjc/s10052-020-08757-3}{\emph{Eur. Phys. J. C}
  {\bfseries 80} (2020) 1173}
  [\href{https://arxiv.org/abs/2007.00507}{{\ttfamily 2007.00507}}].

\bibitem{Chen:2024xwt}
W.~Chen, \emph{{Semi-automatic Calculations of Multi-loop Feynman Amplitudes
  with AmpRed}},  \href{https://arxiv.org/abs/2408.06426}{{\ttfamily
  2408.06426}}.

\bibitem{Artico:2023jrc}
D.~Artico and L.~Magnea, \emph{{Integration-by-parts identities and
  differential equations for parametrised Feynman integrals}},
  \href{https://doi.org/10.1007/JHEP03(2024)096}{\emph{JHEP} {\bfseries 03}
  (2024) 096} [\href{https://arxiv.org/abs/2310.03939}{{\ttfamily
  2310.03939}}].

\bibitem{Wu:2023upw}
Z.~Wu, J.~Boehm, R.~Ma, H.~Xu and Y.~Zhang, \emph{{NeatIBP 1.0, a package
  generating small-size integration-by-parts relations for Feynman integrals}},
  \href{https://doi.org/10.1016/j.cpc.2023.108999}{\emph{Comput. Phys. Commun.}
  {\bfseries 295} (2024) 108999}
  [\href{https://arxiv.org/abs/2305.08783}{{\ttfamily 2305.08783}}].

\bibitem{Guan:2024byi}
X.~Guan, X.~Liu, Y.-Q.~Ma and W.-H.~Wu, \emph{{Blade: A package for
  block-triangular form improved Feynman integrals decomposition}},
  \href{https://arxiv.org/abs/2405.14621}{{\ttfamily 2405.14621}}.

\bibitem{Passarino:1978jh}
G.~Passarino and M.J.G.~Veltman, \emph{{One Loop Corrections for e+ e-
  Annihilation Into mu+ mu- in the Weinberg Model}},
  \href{https://doi.org/10.1016/0550-3213(79)90234-7}{\emph{Nucl. Phys. B}
  {\bfseries 160} (1979) 151}.

\bibitem{Feng:2021enk}
B.~Feng, T.~Li and X.~Li, \emph{{Analytic tadpole coefficients of one-loop
  integrals}}, \href{https://doi.org/10.1007/JHEP09(2021)081}{\emph{JHEP}
  {\bfseries 09} (2021) 081}
  [\href{https://arxiv.org/abs/2107.03744}{{\ttfamily 2107.03744}}].

\bibitem{Hu:2021nia}
C.~Hu, T.~Li and X.~Li, \emph{{One-loop Feynman integral reduction by
  differential operators}},
  \href{https://doi.org/10.1103/PhysRevD.104.116014}{\emph{Phys. Rev. D}
  {\bfseries 104} (2021) 116014}
  [\href{https://arxiv.org/abs/2108.00772}{{\ttfamily 2108.00772}}].

\bibitem{Feng:2022rwj}
B.~Feng, J.~Gong and T.~Li, \emph{{Universal treatment of the reduction for
  one-loop integrals in a projective space}},
  \href{https://doi.org/10.1103/PhysRevD.106.056025}{\emph{Phys. Rev. D}
  {\bfseries 106} (2022) 056025}
  [\href{https://arxiv.org/abs/2204.03190}{{\ttfamily 2204.03190}}].

\bibitem{Feng:2022uqp}
B.~Feng, T.~Li, H.~Wang and Y.~Zhang, \emph{{Reduction of general one-loop
  integrals using auxiliary vector}},
  \href{https://doi.org/10.1007/JHEP05(2022)065}{\emph{JHEP} {\bfseries 05}
  (2022) 065} [\href{https://arxiv.org/abs/2203.14449}{{\ttfamily
  2203.14449}}].

\bibitem{Feng:2022iuc}
B.~Feng and T.~Li, \emph{{PV-reduction of sunset topology with auxiliary
  vector}}, \href{https://doi.org/10.1088/1572-9494/ac7f97}{\emph{Commun.
  Theor. Phys.} {\bfseries 74} (2022) 095201}
  [\href{https://arxiv.org/abs/2203.16881}{{\ttfamily 2203.16881}}].

\bibitem{Feng:2022rfz}
B.~Feng, C.~Hu, T.~Li and Y.~Song, \emph{{Reduction with degenerate Gram matrix
  for one-loop integrals}},
  \href{https://doi.org/10.1007/JHEP08(2022)110}{\emph{JHEP} {\bfseries 08}
  (2022) 110} [\href{https://arxiv.org/abs/2205.03000}{{\ttfamily
  2205.03000}}].

\bibitem{Li:2022cbx}
T.~Li, \emph{{Nontrivial one-loop recursive reduction relation}},
  \href{https://doi.org/10.1007/JHEP07(2023)051}{\emph{JHEP} {\bfseries 07}
  (2023) 051} [\href{https://arxiv.org/abs/2209.11428}{{\ttfamily
  2209.11428}}].

\bibitem{Ossola:2006us}
G.~Ossola, C.G.~Papadopoulos and R.~Pittau, \emph{{Reducing full one-loop
  amplitudes to scalar integrals at the integrand level}},
  \href{https://doi.org/10.1016/j.nuclphysb.2006.11.012}{\emph{Nucl. Phys. B}
  {\bfseries 763} (2007) 147}
  [\href{https://arxiv.org/abs/hep-ph/0609007}{{\ttfamily hep-ph/0609007}}].

\bibitem{Ossola:2007bb}
G.~Ossola, C.G.~Papadopoulos and R.~Pittau, \emph{{Numerical evaluation of
  six-photon amplitudes}},
  \href{https://doi.org/10.1088/1126-6708/2007/07/085}{\emph{JHEP} {\bfseries
  07} (2007) 085} [\href{https://arxiv.org/abs/0704.1271}{{\ttfamily
  0704.1271}}].

\bibitem{Ellis:2007br}
R.K.~Ellis, W.T.~Giele and Z.~Kunszt, \emph{{A Numerical Unitarity Formalism
  for Evaluating One-Loop Amplitudes}},
  \href{https://doi.org/10.1088/1126-6708/2008/03/003}{\emph{JHEP} {\bfseries
  03} (2008) 003} [\href{https://arxiv.org/abs/0708.2398}{{\ttfamily
  0708.2398}}].

\bibitem{Bern:1994zx}
Z.~Bern, L.J.~Dixon, D.C.~Dunbar and D.A.~Kosower, \emph{{One loop n point
  gauge theory amplitudes, unitarity and collinear limits}},
  \href{https://doi.org/10.1016/0550-3213(94)90179-1}{\emph{Nucl. Phys. B}
  {\bfseries 425} (1994) 217}
  [\href{https://arxiv.org/abs/hep-ph/9403226}{{\ttfamily hep-ph/9403226}}].

\bibitem{Bern:1994cg}
Z.~Bern, L.J.~Dixon, D.C.~Dunbar and D.A.~Kosower, \emph{{Fusing gauge theory
  tree amplitudes into loop amplitudes}},
  \href{https://doi.org/10.1016/0550-3213(94)00488-Z}{\emph{Nucl. Phys. B}
  {\bfseries 435} (1995) 59}
  [\href{https://arxiv.org/abs/hep-ph/9409265}{{\ttfamily hep-ph/9409265}}].

\bibitem{Bern:1997sc}
Z.~Bern, L.J.~Dixon and D.A.~Kosower, \emph{{One loop amplitudes for e+ e- to
  four partons}},
  \href{https://doi.org/10.1016/S0550-3213(97)00703-7}{\emph{Nucl. Phys. B}
  {\bfseries 513} (1998) 3}
  [\href{https://arxiv.org/abs/hep-ph/9708239}{{\ttfamily hep-ph/9708239}}].

\bibitem{Britto:2004nc}
R.~Britto, F.~Cachazo and B.~Feng, \emph{{Generalized unitarity and one-loop
  amplitudes in N=4 super-Yang-Mills}},
  \href{https://doi.org/10.1016/j.nuclphysb.2005.07.014}{\emph{Nucl. Phys. B}
  {\bfseries 725} (2005) 275}
  [\href{https://arxiv.org/abs/hep-th/0412103}{{\ttfamily hep-th/0412103}}].

\bibitem{Britto:2005ha}
R.~Britto, E.~Buchbinder, F.~Cachazo and B.~Feng, \emph{{One-loop amplitudes of
  gluons in SQCD}},
  \href{https://doi.org/10.1103/PhysRevD.72.065012}{\emph{Phys. Rev. D}
  {\bfseries 72} (2005) 065012}
  [\href{https://arxiv.org/abs/hep-ph/0503132}{{\ttfamily hep-ph/0503132}}].

\bibitem{Britto:2006sj}
R.~Britto, B.~Feng and P.~Mastrolia, \emph{{The Cut-constructible part of QCD
  amplitudes}}, \href{https://doi.org/10.1103/PhysRevD.73.105004}{\emph{Phys.
  Rev. D} {\bfseries 73} (2006) 105004}
  [\href{https://arxiv.org/abs/hep-ph/0602178}{{\ttfamily hep-ph/0602178}}].

\bibitem{Anastasiou:2006jv}
C.~Anastasiou, R.~Britto, B.~Feng, Z.~Kunszt and P.~Mastrolia,
  \emph{{D-dimensional unitarity cut method}},
  \href{https://doi.org/10.1016/j.physletb.2006.12.022}{\emph{Phys. Lett. B}
  {\bfseries 645} (2007) 213}
  [\href{https://arxiv.org/abs/hep-ph/0609191}{{\ttfamily hep-ph/0609191}}].

\bibitem{Anastasiou:2006gt}
C.~Anastasiou, R.~Britto, B.~Feng, Z.~Kunszt and P.~Mastrolia, \emph{{Unitarity
  cuts and Reduction to master integrals in d dimensions for one-loop
  amplitudes}},
  \href{https://doi.org/10.1088/1126-6708/2007/03/111}{\emph{JHEP} {\bfseries
  03} (2007) 111} [\href{https://arxiv.org/abs/hep-ph/0612277}{{\ttfamily
  hep-ph/0612277}}].

\bibitem{Britto:2006fc}
R.~Britto and B.~Feng, \emph{{Unitarity cuts with massive propagators and
  algebraic expressions for coefficients}},
  \href{https://doi.org/10.1103/PhysRevD.75.105006}{\emph{Phys. Rev. D}
  {\bfseries 75} (2007) 105006}
  [\href{https://arxiv.org/abs/hep-ph/0612089}{{\ttfamily hep-ph/0612089}}].

\bibitem{Britto:2007tt}
R.~Britto and B.~Feng, \emph{{Integral coefficients for one-loop amplitudes}},
  \href{https://doi.org/10.1088/1126-6708/2008/02/095}{\emph{JHEP} {\bfseries
  02} (2008) 095} [\href{https://arxiv.org/abs/0711.4284}{{\ttfamily
  0711.4284}}].

\bibitem{Britto:2010um}
R.~Britto and E.~Mirabella, \emph{{Single Cut Integration}},
  \href{https://doi.org/10.1007/JHEP01(2011)135}{\emph{JHEP} {\bfseries 01}
  (2011) 135} [\href{https://arxiv.org/abs/1011.2344}{{\ttfamily 1011.2344}}].

\bibitem{Feng:2022hyg}
B.~Feng, \emph{{Generation function for one-loop tensor reduction}},
  \href{https://doi.org/10.1088/1572-9494/aca253}{\emph{Commun. Theor. Phys.}
  {\bfseries 75} (2023) 025203}
  [\href{https://arxiv.org/abs/2209.09517}{{\ttfamily 2209.09517}}].

\bibitem{Guan:2023avw}
X.~Guan, X.~Li and Y.-Q.~Ma, \emph{{Exploring the linear space of Feynman
  integrals via generating functions}},
  \href{https://doi.org/10.1103/PhysRevD.108.034027}{\emph{Phys. Rev. D}
  {\bfseries 108} (2023) 034027}
  [\href{https://arxiv.org/abs/2306.02927}{{\ttfamily 2306.02927}}].

\bibitem{Hu:2023mgc}
C.~Hu, T.~Li, J.~Shen and Y.~Xu, \emph{{An explicit expression of generating
  function for one-loop tensor reduction}},
  \href{https://doi.org/10.1007/JHEP02(2024)158}{\emph{JHEP} {\bfseries 02}
  (2024) 158} [\href{https://arxiv.org/abs/2308.13336}{{\ttfamily
  2308.13336}}].

\bibitem{Li:2024rvo}
T.~Li, Y.~Song and L.~Zhang, \emph{{Solving arbitrary one-loop reduction via
  generating function}},  \href{https://arxiv.org/abs/2404.04644}{{\ttfamily
  2404.04644}}.

\bibitem{Mizera:2017rqa}
S.~Mizera, \emph{{Scattering Amplitudes from Intersection Theory}},
  \href{https://doi.org/10.1103/PhysRevLett.120.141602}{\emph{Phys. Rev. Lett.}
  {\bfseries 120} (2018) 141602}
  [\href{https://arxiv.org/abs/1711.00469}{{\ttfamily 1711.00469}}].

\bibitem{Mastrolia:2018uzb}
P.~Mastrolia and S.~Mizera, \emph{{Feynman Integrals and Intersection Theory}},
  \href{https://doi.org/10.1007/JHEP02(2019)139}{\emph{JHEP} {\bfseries 02}
  (2019) 139} [\href{https://arxiv.org/abs/1810.03818}{{\ttfamily
  1810.03818}}].

\bibitem{Frellesvig:2019kgj}
H.~Frellesvig, F.~Gasparotto, S.~Laporta, M.K.~Mandal, P.~Mastrolia,
  L.~Mattiazzi et~al., \emph{{Decomposition of Feynman Integrals on the Maximal
  Cut by Intersection Numbers}},
  \href{https://doi.org/10.1007/JHEP05(2019)153}{\emph{JHEP} {\bfseries 05}
  (2019) 153} [\href{https://arxiv.org/abs/1901.11510}{{\ttfamily
  1901.11510}}].

\bibitem{Frellesvig:2019uqt}
H.~Frellesvig, F.~Gasparotto, M.K.~Mandal, P.~Mastrolia, L.~Mattiazzi and
  S.~Mizera, \emph{{Vector Space of Feynman Integrals and Multivariate
  Intersection Numbers}},
  \href{https://doi.org/10.1103/PhysRevLett.123.201602}{\emph{Phys. Rev. Lett.}
  {\bfseries 123} (2019) 201602}
  [\href{https://arxiv.org/abs/1907.02000}{{\ttfamily 1907.02000}}].

\bibitem{Mizera:2019vvs}
S.~Mizera and A.~Pokraka, \emph{{From Infinity to Four Dimensions: Higher
  Residue Pairings and Feynman Integrals}},
  \href{https://doi.org/10.1007/JHEP02(2020)159}{\emph{JHEP} {\bfseries 02}
  (2020) 159} [\href{https://arxiv.org/abs/1910.11852}{{\ttfamily
  1910.11852}}].

\bibitem{Mizera:2020wdt}
S.~Mizera, \emph{{Status of Intersection Theory and Feynman Integrals}},
  \href{https://doi.org/10.22323/1.383.0016}{\emph{PoS} {\bfseries MA2019}
  (2019) 016} [\href{https://arxiv.org/abs/2002.10476}{{\ttfamily
  2002.10476}}].

\bibitem{Frellesvig:2020qot}
H.~Frellesvig, F.~Gasparotto, S.~Laporta, M.K.~Mandal, P.~Mastrolia,
  L.~Mattiazzi et~al., \emph{{Decomposition of Feynman Integrals by
  Multivariate Intersection Numbers}},
  \href{https://doi.org/10.1007/JHEP03(2021)027}{\emph{JHEP} {\bfseries 03}
  (2021) 027} [\href{https://arxiv.org/abs/2008.04823}{{\ttfamily
  2008.04823}}].

\bibitem{Caron-Huot:2021xqj}
S.~Caron-Huot and A.~Pokraka, \emph{{Duals of Feynman integrals. Part I.
  Differential equations}},
  \href{https://doi.org/10.1007/JHEP12(2021)045}{\emph{JHEP} {\bfseries 12}
  (2021) 045} [\href{https://arxiv.org/abs/2104.06898}{{\ttfamily
  2104.06898}}].

\bibitem{Caron-Huot:2021iev}
S.~Caron-Huot and A.~Pokraka, \emph{{Duals of Feynman Integrals. Part II.
  Generalized unitarity}},
  \href{https://doi.org/10.1007/JHEP04(2022)078}{\emph{JHEP} {\bfseries 04}
  (2022) 078} [\href{https://arxiv.org/abs/2112.00055}{{\ttfamily
  2112.00055}}].

\bibitem{Chestnov:2022alh}
V.~Chestnov, F.~Gasparotto, M.K.~Mandal, P.~Mastrolia, S.J.~Matsubara-Heo,
  H.J.~Munch et~al., \emph{{Macaulay matrix for Feynman integrals: linear
  relations and intersection numbers}},
  \href{https://doi.org/10.1007/JHEP09(2022)187}{\emph{JHEP} {\bfseries 09}
  (2022) 187} [\href{https://arxiv.org/abs/2204.12983}{{\ttfamily
  2204.12983}}].

\bibitem{Fontana:2023amt}
G.~Fontana and T.~Peraro, \emph{{Reduction to master integrals via intersection
  numbers and polynomial expansions}},
  \href{https://doi.org/10.1007/JHEP08(2023)175}{\emph{JHEP} {\bfseries 08}
  (2023) 175} [\href{https://arxiv.org/abs/2304.14336}{{\ttfamily
  2304.14336}}].

\bibitem{Brunello:2023rpq}
G.~Brunello, V.~Chestnov, G.~Crisanti, H.~Frellesvig, M.K.~Mandal and
  P.~Mastrolia, \emph{{Intersection numbers, polynomial division and relative
  cohomology}}, \href{https://doi.org/10.1007/JHEP09(2024)015}{\emph{JHEP}
  {\bfseries 09} (2024) 015}
  [\href{https://arxiv.org/abs/2401.01897}{{\ttfamily 2401.01897}}].

\bibitem{Baikov:1996iu}
P.A.~Baikov, \emph{{Explicit solutions of the multiloop integral recurrence
  relations and its application}},
  \href{https://doi.org/10.1016/S0168-9002(97)00126-5}{\emph{Nucl. Instrum.
  Meth. A} {\bfseries 389} (1997) 347}
  [\href{https://arxiv.org/abs/hep-ph/9611449}{{\ttfamily hep-ph/9611449}}].

\bibitem{Lee:2010wea}
R.N.~Lee, \emph{{Calculating multiloop integrals using dimensional recurrence
  relation and $D$-analyticity}},
  \href{https://doi.org/10.1016/j.nuclphysbps.2010.08.032}{\emph{Nucl. Phys. B
  Proc. Suppl.} {\bfseries 205-206} (2010) 135}
  [\href{https://arxiv.org/abs/1007.2256}{{\ttfamily 1007.2256}}].

\bibitem{Lu:2024dsb}
M.~Lu, Z.~Wang and L.L.~Yang, \emph{{Intersection theory, relative cohomology
  and the Feynman parametrization}},
  \href{https://arxiv.org/abs/2411.05226}{{\ttfamily 2411.05226}}.

\bibitem{Cheng:1969ab}
H.~Cheng and T.T.~Wu, \emph{{High-energy collision processes in quantum
  electrodynamics. iii}},
  \href{https://doi.org/10.1103/PhysRev.182.1873}{\emph{Phys. Rev.} {\bfseries
  182} (1969) 1873}.

\bibitem{Cheng:1987ga}
H.~Cheng and T.T.~Wu, \emph{{EXPANDING PROTONS: SCATTERING AT HIGH-ENERGIES}}
  (1987).

\bibitem{Smirnov:2012gma}
V.A.~Smirnov, \emph{{Analytic tools for Feynman integrals}}, vol.~250 (2012),
  \href{https://doi.org/10.1007/978-3-642-34886-0}{10.1007/978-3-642-34886-0}.

\bibitem{Binoth:2005ff}
T.~Binoth, J.P.~Guillet, G.~Heinrich, E.~Pilon and C.~Schubert, \emph{{An
  Algebraic/numerical formalism for one-loop multi-leg amplitudes}},
  \href{https://doi.org/10.1088/1126-6708/2005/10/015}{\emph{JHEP} {\bfseries
  10} (2005) 015} [\href{https://arxiv.org/abs/hep-ph/0504267}{{\ttfamily
  hep-ph/0504267}}].

\bibitem{Britto:2023rig}
R.~Britto, \emph{{Generalized Cuts of Feynman Integrals in Parameter Space}},
  \href{https://doi.org/10.1103/PhysRevLett.131.091601}{\emph{Phys. Rev. Lett.}
  {\bfseries 131} (2023) 091601}
  [\href{https://arxiv.org/abs/2305.15369}{{\ttfamily 2305.15369}}].

\bibitem{Lee:2013hzt}
R.N.~Lee and A.A.~Pomeransky, \emph{{Critical points and number of master
  integrals}}, \href{https://doi.org/10.1007/JHEP11(2013)165}{\emph{JHEP}
  {\bfseries 11} (2013) 165} [\href{https://arxiv.org/abs/1308.6676}{{\ttfamily
  1308.6676}}].

\bibitem{Arkani-Hamed:2022cqe}
N.~Arkani-Hamed, A.~Hillman and S.~Mizera, \emph{{Feynman polytopes and the
  tropical geometry of UV and IR divergences}},
  \href{https://doi.org/10.1103/PhysRevD.105.125013}{\emph{Phys. Rev. D}
  {\bfseries 105} (2022) 125013}
  [\href{https://arxiv.org/abs/2202.12296}{{\ttfamily 2202.12296}}].

\bibitem{Klausen:2023gui}
R.P.~Klausen, \emph{{Hypergeometric feynman integrals}}, Ph.D. thesis, Mainz
  U., Mainz U., 2023.
\newblock \href{https://arxiv.org/abs/2302.13184}{{\ttfamily 2302.13184}}.
\newblock 10.25358/openscience-8527.

\bibitem{Weinzierl:2022eaz}
S.~Weinzierl, \emph{{Feynman Integrals. A Comprehensive Treatment for Students
  and Researchers}}, UNITEXT for Physics, Springer (2022),
  \href{https://doi.org/10.1007/978-3-030-99558-4}{10.1007/978-3-030-99558-4},
  [\href{https://arxiv.org/abs/2201.03593}{{\ttfamily 2201.03593}}].

\bibitem{Aomoto:1414035}
K.~Aomoto and M.~Kita, \emph{{Theory of Hypergeometric Functions}}, Springer
  Monographs in Mathematics, Springer (2011),
  \href{https://doi.org/10.1007/978-4-431-53938-4}{10.1007/978-4-431-53938-4}.

\bibitem{zbMATH00713739}
K.~Matsumoto, \emph{Quadratic identities for hypergeometric series of type
  {{\((k,l)\)}}}, \href{https://doi.org/10.2206/kyushujm.48.335}{\emph{Kyushu
  J. Math.} {\bfseries 48} (1994) 335}.

\bibitem{zbMATH01270294}
K.~Matsumoto, \emph{Intersection numbers for logarithmic {{\(K\)}}-forms},
  {\emph{Osaka J. Math.} {\bfseries 35} (1998) 873}.

\bibitem{zbMATH02112802}
K.~Ohara, Y.~Sugiki and N.~Takayama, \emph{Quadratic relations for generalized
  hypergeometric functions {{\(_ pF_{p-1}\)}}.},
  \href{https://doi.org/10.1619/fesi.46.213}{\emph{Funkc. Ekvacioj, Ser. Int.}
  {\bfseries 46} (2003) 213}.

\bibitem{zbMATH06267077}
Y.~Goto, \emph{Twisted cycles and twisted period relations for {Lauricella}'s
  hypergeometric function {{\(F_{c}\)}}},
  \href{https://doi.org/10.1142/S0129167X13500948}{\emph{Int. J. Math.}
  {\bfseries 24} (2013) 19}.

\bibitem{zbMATH06447521}
Y.~Goto and K.~Matsumoto, \emph{The monodromy representation and twisted period
  relations for {Appell}'s hypergeometric function {{\(F_4\)}}},
  \href{https://doi.org/10.1215/00277630-2873714}{\emph{Nagoya Math. J.}
  {\bfseries 217} (2015) 61}.

\bibitem{zbMATH06502598}
Y.~Goto, \emph{Twisted period relations for {Lauricella}'s hypergeometric
  functions {{\(F_A\)}}}, {\emph{Osaka J. Math.} {\bfseries 52} (2015) 861}.

\bibitem{zbMATH06454357}
Y.~Goto, \emph{Intersection numbers and twisted period relations for the
  generalized hypergeometric function {{\(_{m+1}F_{m}\)}}},
  \href{https://doi.org/10.2206/kyushujm.69.203}{\emph{Kyushu J. Math.}
  {\bfseries 69} (2015) 203}.

\bibitem{zbMATH07531013}
S.-J.~Matsubara-Heo and N.~Takayama, \emph{An algorithm of computing cohomology
  intersection number of hypergeometric integrals},
  \href{https://doi.org/10.1017/nmj.2021.2}{\emph{Nagoya Math. J.} {\bfseries
  246} (2022) 256}.

\bibitem{zbMATH07527773}
Y.~Goto and S.-J.~Matsubara-Heo, \emph{Homology and cohomology intersection
  numbers of {GKZ} systems},
  \href{https://doi.org/10.1016/j.indag.2021.12.002}{\emph{Indag. Math., New
  Ser.} {\bfseries 33} (2022) 546}.

\bibitem{zbMATH07733418}
S.-J.~Matsubara-Heo, \emph{Localization formulas of cohomology intersection
  numbers}, \href{https://doi.org/10.2969/jmsj/87738773}{\emph{J. Math. Soc.
  Japan} {\bfseries 75} (2023) 909}.

\bibitem{Denner:2005nn}
A.~Denner and S.~Dittmaier, \emph{{Reduction schemes for one-loop tensor
  integrals}},
  \href{https://doi.org/10.1016/j.nuclphysb.2005.11.007}{\emph{Nucl. Phys. B}
  {\bfseries 734} (2006) 62}
  [\href{https://arxiv.org/abs/hep-ph/0509141}{{\ttfamily hep-ph/0509141}}].

\bibitem{Tarasov:1996br}
O.V.~Tarasov, \emph{{Connection between Feynman integrals having different
  values of the space-time dimension}},
  \href{https://doi.org/10.1103/PhysRevD.54.6479}{\emph{Phys. Rev. D}
  {\bfseries 54} (1996) 6479}
  [\href{https://arxiv.org/abs/hep-th/9606018}{{\ttfamily hep-th/9606018}}].

\bibitem{Tarasov:1997kx}
O.V.~Tarasov, \emph{{Generalized recurrence relations for two loop propagator
  integrals with arbitrary masses}},
  \href{https://doi.org/10.1016/S0550-3213(97)00376-3}{\emph{Nucl. Phys. B}
  {\bfseries 502} (1997) 455}
  [\href{https://arxiv.org/abs/hep-ph/9703319}{{\ttfamily hep-ph/9703319}}].

\bibitem{Lee:2009dh}
R.N.~Lee, \emph{{Space-time dimensionality D as complex variable: Calculating
  loop integrals using dimensional recurrence relation and analytical
  properties with respect to D}},
  \href{https://doi.org/10.1016/j.nuclphysb.2009.12.025}{\emph{Nucl. Phys. B}
  {\bfseries 830} (2010) 474}
  [\href{https://arxiv.org/abs/0911.0252}{{\ttfamily 0911.0252}}].

\bibitem{Lee:2010ug}
R.N.~Lee, A.V.~Smirnov and V.A.~Smirnov, \emph{{Dimensional recurrence
  relations: an easy way to evaluate higher orders of expansion in
  $\epsilon$}},
  \href{https://doi.org/10.1016/j.nuclphysbps.2010.09.011}{\emph{Nucl. Phys. B
  Proc. Suppl.} {\bfseries 205-206} (2010) 308}
  [\href{https://arxiv.org/abs/1005.0362}{{\ttfamily 1005.0362}}].

\bibitem{Frellesvig:2018ymi}
H.A.~Frellesvig, R.~Bonciani, V.~Del~Duca, F.~Moriello, J.~Henn and V.~Smirnov,
  \emph{{Non-planar two-loop Feynman integrals contributing to Higgs plus jet
  production}}, \href{https://doi.org/10.22323/1.303.0076}{\emph{PoS}
  {\bfseries LL2018} (2018) 076}.

\bibitem{Landau:1959fi}
L.D.~Landau, \emph{{On the Analytic Properties of Vertex Parts in Quantum Field
  Theory}}, \href{https://doi.org/10.1016/B978-0-08-010586-4.50103-6}{\emph{Zh.
  Eksp. Teor. Fiz.} {\bfseries 37} (1960) 62}.

\bibitem{Coro:2025kha}
F.~Coro, P.P.~Novichkov, B.~Page and Q.~Song, \emph{{Feynman Integral Reduction
  and Landau Singularities}},
  \href{https://arxiv.org/abs/2512.05869}{{\ttfamily 2512.05869}}.

\bibitem{Jiang:2024eaj}
X.~Jiang, J.~Liu, X.~Xu and L.L.~Yang, \emph{{Symbol letters of Feynman
  integrals from Gram determinants}},
  \href{https://doi.org/10.1016/j.physletb.2025.139443}{\emph{Phys. Lett. B}
  {\bfseries 864} (2025) 139443}
  [\href{https://arxiv.org/abs/2401.07632}{{\ttfamily 2401.07632}}].

\bibitem{Correia:2025wtb}
M.~Correia, M.~Giroux and S.~Mizera, \emph{{SOFIA: Singularities of Feynman
  integrals automatized}},
  \href{https://doi.org/10.1016/j.cpc.2025.109970}{\emph{Comput. Phys. Commun.}
  {\bfseries 320} (2026) 109970}
  [\href{https://arxiv.org/abs/2503.16601}{{\ttfamily 2503.16601}}].

\end{thebibliography}\endgroup

\end{document}